\documentclass[%
reprint,
superscriptaddress,
 amsmath,amssymb,
 aps,
]{revtex4-2}

\usepackage{graphicx}
\usepackage{dcolumn}
\usepackage{bm}
\usepackage{hyperref}

\usepackage{multirow}
\usepackage{comment}
\usepackage{chngcntr}

\begin{document}


\title{Improving the Predictive Capability of the Fission Reaction Event Yield Algorithm ($\mathtt{FREYA}$)}

\author{A.~E. Tuckey}
\email{atuckey@umich.edu}
\affiliation{Department of Nuclear Engineering and Radiological Sciences, University of Michigan, Ann Arbor, MI 48109, USA}
\author{R. Vogt}
\affiliation{Nuclear and Chemical Sciences Division, Lawrence Livermore National Laboratory, Livermore, CA 94550, USA}
\affiliation{Department of Physics and Astronomy, University of California, Davis, CA 95616, USA}
\author{D. Breitenmoser}
\affiliation{Department of Nuclear Engineering and Radiological Sciences, University of Michigan, Ann Arbor, MI 48109, USA}
\author{S.~D. Clarke}
\affiliation{Department of Nuclear Engineering and Radiological Sciences, University of Michigan, Ann Arbor, MI 48109, USA}
\author{S.~A. Pozzi}
\affiliation{Department of Nuclear Engineering and Radiological Sciences, University of Michigan, Ann Arbor, MI 48109, USA}
\author{M. Devlin}
\affiliation{Physics Division, Los Alamos National Laboratory, Los Alamos, NM 87545, USA}

\date{\today}

\begin{abstract}
We determine the optimal parameters for thermal neutron-induced fission of $^{233,235}$U and $^{239,241}$Pu in the complete event fission model $\mathtt{FREYA}$. First, we revisit and, in most cases, improve the prior values determined for spontaneous fission using a genetic algorithm.  Our optimization procedure is then applied to thermal neutron-induced fission, replacing the empirically-chosen parameters previously employed in $\mathtt{FREYA}$. Finally, the optimized parameter values are used to make predictions for spontaneous and thermal neutron-induced fission observables not included in the fits. This work represents the first step in a broader program to study neutron-induced fission as a function of incident neutron energy, to systematically improve the performance of $\mathtt{FREYA}$, capturing any energy-dependent trends of the physics-based $\mathtt{FREYA}$ parameters.
\end{abstract}

\maketitle


\section{Introduction}
\label{sec:introduction}

Despite being discovered over eighty years ago, nuclear fission has yet to be fully described within a theoretical framework. Phenomenological models such as the Fission Reaction Event Yield Algorithm ($\mathtt{FREYA}$)~\cite{Verbeke2018} help address knowledge gaps in the fission process by providing a predictive capability when no experimental data exist. Rather than treating fission as a black box, codes like $\mathtt{FREYA}$ simulate the complete event, conserving energy and momentum and thus allowing the correlations among all parts of the fission process to be addressed. $\mathtt{FREYA}$ is flexible, making it possible to address multiple fission channels based on the excitation of the compound nucleus.  It is also sufficiently fast to be included as an option in large neutron transport codes without a significant reduction in processing time~\cite{Talou2018}.  Thus, $\mathtt{FREYA}$ is a useful tool for understanding correlated fission data and for identifying additional fission measurements to inform theory and model development.

The events generated by $\mathtt{FREYA}$ depend on five physics-based parameters. The results for a given parameter set can be compared to existing experimental data and evaluations to determine how effectively each set describes the data. Determining the optimal sets of these parameters for each isotope and, for induced fission, a given incident particle energy, improves the accuracy of the $\mathtt{FREYA}$ simulations and enhances the reliability of its predictions when experimental data are unavailable. Van Dyke {\it et al.}~\cite{VanDyke2019} determined the optimal parameter values for spontaneous fission based on simulated annealing \cite{Kirkpatrick1983}. In the most recent release of $\mathtt{FREYA}$~\cite{Verbeke2018}, the parameters for neutron-induced fission were tuned by hand or, if no data were available, assumed to be energy and isotope independent.

Here we adopt a genetic algorithm and confirm that this approach agrees with the simulated annealing results~\cite{VanDyke2019} for spontaneous fission of $^{252}$Cf when the same data are employed in the fit. We next use the genetic algorithm to update and, typically, significantly improve the parameter values for spontaneous fission relative to the results of Ref.~\cite{VanDyke2019}.  We then apply the optimization procedure developed for spontaneous fission to determine the optimal parameter values for thermal neutron-induced fission of $^{233,235}$U and $^{239,241}$Pu. This work is the first to systematically study the isotope-dependent trend of the $\mathtt{FREYA}$ parameters for thermal neutron-induced fission and test them against previous assumptions.  It also represents a first step in the exploration of the energy dependence of the inputs for neutron-induced fission. We perform a full statistical analysis, including variances and covariances, and compare both to data and results employing the previous default $\mathtt{FREYA}$ parameters.

Section \ref{sec:parameters} describes the $\mathtt{FREYA}$ parameters and their physical origin. Section \ref{sec:methods} discusses the numerical methods used to perform the parameter determinations, while Section \ref{sec:data} describes the data used in the fits. Section \ref{sec:results} presents the optimized parameters and compares calculations employing these results to data and previous $\mathtt{FREYA}$ calculations. In Section \ref{sec:conclusion} we conclude. Additional results can be found in Appendix~\ref{sec:A}, while a more detailed discussion of the genetic algorithm hyperparameters can be found in Appendix~\ref{sec:B}.

\section{$\mathtt{FREYA}$ parameter description}
\label{sec:parameters}

We now provide a brief description of nuclear fission as implemented in $\mathtt{FREYA}$ and identify the five physics-based parameters that the model requires. 

Neutron-induced fission begins with a target nucleus of mass $A$ absorbing an incident neutron. The resulting compound nucleus (CN) of mass $A_{\rm CN} = A + 1$ may emit one or more neutrons prior to fission, in which case the daughter nucleus has mass $A_0 = A_{\rm CN} - n$, where $n$ is the number of pre-scission neutrons emitted. In thermal neutron-induced fission, the excitation energy of the compound nucleus is below the energy threshold for neutron emission. In the case of spontaneous fission, $A_{\rm CN} = A$.

Fission can occur if the excitation energy of the daughter nucleus following pre-scission emission, if applicable, is above the fission barrier. The daughter nucleus splits into two fragment nuclei, typically one light and one heavy, which we denote by $A_L$ and $A_H$, respectively. (The possibility of ternary fission is neglected in $\mathtt{FREYA}$.)  The fragment yields as a function of the fragment mass, as well as the average total kinetic energy, $\overline{{\rm TKE}}$, of the fragments as a function of the heavy fragment mass, $A_H$, are sampled from data. If no data exist, modeled yields may be used instead.  Although the mass and charge yields may be modeled successfully in several theoretical approaches, TKE$(A_H)$ is more difficult, and no successful models exist so far (see {\it e.g.}\ Ref.~\cite{Bjelcic:2025dwt} for a recent attempt).

The total energy released by scission (the $Q$ value) is calculated from the mass difference between the fissioning nucleus and the fragments, $Q = M_0c^2 - M_Lc^2 - M_Hc^2$. Given the $Q$ value and the sampled $\overline{{\rm TKE}}$, the total excitation energy at scission, $E_{\rm sc}^*$, is determined by energy conservation,
\begin{equation}
\label{eq:scission}
    E_{\rm sc}^* = Q - \overline{{\rm TKE}} = E_{\rm stat} + E_{\rm rot} \, \, .
\end{equation}
Here $E_{\rm sc}^*$ includes both the statistical excitation energy, $E_{\rm stat}$, and the rotational excitation energy, $E_{\rm rot}$, available to the two fragment nuclei at scission. 

The corresponding scission temperature, $T_{\rm sc}$, is obtained from the relation 
\begin{equation}
\label{eq:e0}
    E_{\rm sc}^* = a T_{\rm sc}^2 \, \, ,
\end{equation}
where $a$ is the level density parameter.  $\mathtt{FREYA}$ uses the Ignatyuk parameterization of the level density where, at high energies, $a \sim A_0/e_0$~\cite{Vogt:2011je}. The constant $e_0$ is the first parameter required by $\mathtt{FREYA}$. 

One might expect that $e_0$ is universal, independent of isotope or energy.  Although the value of $e_0$ is related to the Fermi energy of the nucleus~\cite{Bohr1969}, a range of $e_0$ values have been proposed, from 7~MeV$^{-1}$ up to 15~MeV$^{-1}$~\cite{Sepiani_2024}. While the Ignatyuk parameterization used in $\mathtt{FREYA}$ is independent of energy above $\sim50$~MeV/nucleon, at the low excitation energies of spontaneous and thermal neutron-induced fission, some energy dependence remains~\cite{Sepiani_2024}.  Furthermore, the low-energy level densities depend on the level scheme of each nucleus, making the Ignatyuk parameterization somewhat problematic at very low energies because it ignores the level structure in this region.  Thus, assuming a single value of $e_0$ for all fission fragments and all energies is an approximation making it quite conceivable that the optimized value of $e_0$ is not constant, but instead depends on the isotope and the excitation energy.  (Note also that same value of $e_0$ is used throughout $\mathtt{FREYA}$ wherever the temperature is determined, including neutron and photon emission from the fragments.)

The overall rigid rotation of the system prior to scission, induced by the absorption of the incident neutron and the recoil(s) from pre-fission evaporation(s), dictates the mean angular momenta of the two fragments. There are also fluctuations around this value attributed to the wriggling and bending modes that contribute to $E_{\rm rot}$~\cite{Randrup:2014aaa}. The relative degree of these fluctuations is given by
\begin{equation}
\label{eq:cS}
    T_S = c_S T_{\rm sc} \, \, ,
\end{equation}
where $c_S$, the ratio of the spin temperature, $T_S$, to the scission temperature, $T_{\rm sc}$, is the second parameter. Because it is related to fluctuations in the angular momentum of the fragment, $c_S$ directly affects the observed photon emission.  Fluctuations in the values of the fragment spins due to the introduction of $c_S$ are balanced by shifts in the orbital angular momentum to maintain total angular momentum conservation.

We note that the angular momentum fluctuations described here are induced by the exchange of nucleons between the proto-fragments prior to scission; see Ref.~\cite{Randrup:2022} for details and references. In the published version of $\mathtt{FREYA}$, only the wriggling and bending modes, perpendicular to the fission axis, are excited. Reference~\cite{Randrup:2022} addresses how photon observables may change if the twisting mode, parallel to the fission axis, is also excited. In such a case, $c_S$ could be modified to study the excitation of each mode separately, as done in Ref.~\cite{Randrup:2022}. For example, the spin correlations between the fragments change if twisting is introduced. 

Once the rotational excitation energy, $E_{\rm rot}$, has been taken into account, the statistical excitation energy, $E_{\rm stat}$, is determined from Eq.~(\ref{eq:scission}) and is initially partitioned as $E_{\rm stat} = E_L^* + E_H^*$ where the $*$ indicates that the statistical excitation is partitioned according to the level density parameters of the fragments. Thus $E_i^* = E_{\rm stat} \left( \frac{a_i}{a_L + a_H} \right)$ where $i = \{L, H\}$.  Because the observed neutron multiplicities, $\nu$, per fragment group suggest that the light fragment emits more neutrons on average than the heavy fragment for spontaneous and thermal neutron-induced fission~\cite{Nishio:1998gvi,Vorobyev:2004}, the partition is modified via the third parameter, $x$,
\begin{equation}
\label{eq:x}
    \overline E_L^* = xE_L^* \, \, , \quad \overline E_H^* = E_{\rm stat} - \overline E_L^* \, \, ,
\end{equation}
assumed to be larger than unity to ensure that $\nu_L > \nu_H$. 

The $x$ parameter directly affects the fragment mass dependence of the neutron multiplicity, $\nu(A)$, as well as the angular correlations between two neutrons, denoted as $n$--$n$ correlations. The angular correlations arise from prompt neutron emission from the fragments.  While these neutrons are emitted isotropically in the rest frame of the excited fragment, after boosting to the laboratory frame, the neutrons will preferentially move in the direction of fragment travel due to its momentum.  Assuming binary fission, the light and heavy fragments travel in equal and opposite directions to conserve momentum. Therefore, two neutrons emitted from the same fragment will be more aligned than neutrons emitted from opposite fragments. 

Increasing $x$ enhances the statistical excitation energy available to the light fragment, leading to a greater fraction of the prompt neutrons coming from the light fragment. Thus, one effect of increasing the $x$ parameter is to enhance the $n$--$n$ correlation in the direction of the light fragment, at 0~degrees.  (Modifying $x$ will also change the neutron-light fragment angular correlations.) These correlations thus depend on the average neutron multiplicity and the neutron energy.

The effect of thermal fluctuations on the mean fragment excitation energies is considered by sampling an energy fluctuation, $\delta E_i^*$, from a normal distribution with variance
\begin{equation}
\label{eq:c}
\sigma_{\overline E_i^*}^2 = 2c^2 \overline E_i^* T_i \, \, 
\end{equation}
and adjusting the fragment excitation energies accordingly,
\begin{equation}
\label{eq:fluctuation}
    E_i^* = \overline E_i^* + \delta E_i^* \, \, , \quad i=\{L,H\} \, \, .
\end{equation}
The fourth parameter $c$ controls the truncation of the normal distribution at the maximum available excitation. This parameter thus directly affects the neutron multiplicity distribution and its moments.  Energy conservation is maintained by making a compensating opposite fluctuation in the total kinetic energy, ${\rm TKE} = \overline{{\rm TKE}} - \delta E_L^* - \delta E_H^*$.

The fifth and final parameter, $d$TKE, is introduced to shift the average TKE in order to reproduce the measured average neutron multiplicity, $\overline {\nu}$. The sampled $\overline {\rm TKE}(A_H)$ data used to optimize the parameters may have unquantified systematic uncertainties or low statistics.


Two other quantities, $g_{\rm min}$ and $t_{\rm max}$, the minimum photon energy detected and the maximum photon measurement time, respectively, must be set based on a given detector setup.  These parameters are measurement specific and therefore cannot be tuned but can influence the photon observables, particularly $g_{\rm min}$. This parameter is typically $\sim 100$~keV, depending on the type of photon detector, and is necessary to prevent unlimited emission of low energy photons in $\mathtt{FREYA}$.  The photon measurement time typically does not have a strong effect on the photon observables since it is on the order of tens of nanoseconds.  See Ref.~\cite{Vogt:2017} for further details of how these measurement-specific values affect the photon observables.

\section{Computational methods}
\label{sec:methods}

The $\mathtt{FREYA}$ parameters are determined by an iterative $\chi^2$ minimization procedure where the five parameters are randomly selected over a given physical range, as described in detail in this section.  One million $\mathtt{FREYA}$ events are generated for each set of five parameters. The output from the generated events contains the full kinematic information for the fragments and the prompt neutrons and photons. This kinematic information is used to calculate physical observables which are then compared to measured or evaluated data. The quantities extracted for this work are summarized in Table~\ref{tab:observables}.   This list is incomplete because other desired observables can also be extracted, including other correlations beyond the two-neutron angular distribution mentioned in the table.

The moments of the neutron multiplicity distribution, $P(\nu)$, are defined as 
\begin{equation}
\nu_n = \sum_\nu \frac{\nu!}{(\nu - n)!} P(\nu) \, \, .
\end{equation}
These moments numerically encapsulate the shape of the neutron multiplicity distribution.  Note that $\nu_1 \equiv \overline \nu$ in Table~\ref{tab:observables}.

\begin{table}
\caption{Non-exhaustive list of observables that can be extracted from the $\mathtt{FREYA}$ output and, where available, are included in the optimization (unless noted otherwise in Secs.~\ref{sec:data} and~\ref{sec:results}).}
\label{tab:observables}
\begin{ruledtabular}
\begin{tabular}{ll}
Symbol & Description \\
\colrule
$P(\nu)$ & neutron multiplicity distribution \\
$\overline \nu$ & average neutron multiplicity \\
$\nu_2$ & second moment of the neutron multiplicity \\ & distribution \\
$\nu_3$ & third moment of the neutron multiplicity \\ & distribution \\
$\overline \nu(A)$ & average neutron multiplicity as a function of \\ & fragment mass \\
$\overline \nu(\rm TKE)$ & average neutron multiplicity as a function of \\ & fragment TKE \\
PFNS & prompt fission neutron spectrum \\
$n$--$n$ & correlation between two neutron emission angles \\
$P(N_{\gamma})$ & photon multiplicity distribution \\
$\overline N_\gamma$ & average photon multiplicity \\
$\overline N_\gamma(A)$ & average photon multiplicity as a function of \\ & fragment mass \\
$\overline \epsilon_\gamma$ & average photon energy \\
$\overline \epsilon_{\gamma}(A)$ & average photon energy as a function of fragment \\ & mass \\
$\overline E_\gamma (A)$ & average total photon energy as a function of \\ & fragment mass \\
$\overline E_\gamma(\rm TKE)$ & average total photon energy as a function of \\ & fragment TKE \\
PFGS & prompt fission photon (gamma) spectrum \\
\end{tabular}
\end{ruledtabular}
\end{table}

The $\mathtt{FREYA}$ output is compared to the available data for a given fissioning system.  The reduced $\chi^2$, $\chi_O^2$, is calculated for each observable as
\begin{equation}
\label{eq:chisq}
    \chi_O^2 = \frac{1}{n-5} \sum_{i=1}^n \frac{(O_i-E_i)^2}{\sigma_i^2} \, \, ,
\end{equation}
where $i=1,...,n$ runs over the bins of the distribution; $O_i$ is the value of the observable returned by $\mathtt{FREYA}$ in the given bin; $E_i$ is the experimental or evaluated result for the observable in bin $i$; and $\sigma_i$ is the uncertainty on $E_i$. We do not include the statistical uncertainty on $O_i$ resulting from the $\mathtt{FREYA}$ calculation because, in most cases, it is well below the uncertainty on the data employed in our optimization. Exceptions may occur in the high-energy tails of the PFNS or PFGS where the $\mathtt{FREYA}$ result has very low statistics, with only a few events contributing to this part of the spectrum.  These cases do not contribute significantly to the overall $\chi_O^2$. The sum over all bins in Eq.~(\ref{eq:chisq}) is divided by the number of degrees of freedom, $n-5$, to take into account the five fitted parameters. In the case of single-valued observables such as $\overline \nu$, we simply take $\chi_O^2 = (O_i-E_i)^2/\sigma_i^2$. Since some uncertainty is required to calculate $\chi_O^2$, we assume a 5\% uncertainty on data where no uncertainties are reported. The total $\chi^2$ is the sum over all observables with available data, 
\begin{equation}
\label{eq:total_chisq}
    \chi^2 = \sum_O \chi_O^2 \, \, .
\end{equation}
This total $\chi^2$ is treated as the return value of an objective function.

Reference~\cite{VanDyke2019} optimized the spontaneously fissioning nuclei included in $\mathtt{FREYA}$ using a $\chi^2$ minimization based on simulated annealing~\cite{Kirkpatrick1983}. Here we have instead chosen to employ a genetic algorithm~\cite{Goldberg1989} to optimize the $\mathtt{FREYA}$ parameters for spontaneous and thermal neutron-induced fission. Our primary motivation for using a genetic algorithm over alternative optimization methods is its potential to reduce the optimization time when applied more broadly to neutron-induced fission. We discuss this point further later in this section. Along with our new results on thermal neutron-induced fission, we have confirmed and, in many cases, improved the simulated annealing results of Ref.~\cite{VanDyke2019} for spontaneous fission using our genetic algorithm, as detailed in this work. A brief description of our genetic algorithm is now provided. 

Genetic algorithms employ the mechanics of evolution found in nature. To start, a population of size $p$ is randomly initialized and the fitness score, $F$, of each member of the population is evaluated. Individuals in the current population are selected as parents for the next generation based on their level of fitness. $k$ individuals from the population are randomly selected, and the individual with the highest fitness score is passed on to be a parent to the next generation. An individual is not excluded from future selections once it has been sampled. Consequently, an individual can be selected as a parent multiple times. Thus, highly fit individuals have a higher probability of being selected as parents. 

Once $p$ parents have been selected,  crossover and mutation operators are applied to create the children that populate the next generation. A crossover consists of randomly selecting two parents and swapping genes between the two according to a crossover rate, $c$. If a random number is less than $c$, genes are randomly swapped between parents to create two children. Otherwise, the resulting children are identical to their parents. Mutations, which maintain diversity in the population, are introduced by randomly changing the gene sequence according to the mutation rate, $m$. Each gene in the sequence can mutate if a random number selected for that gene is less than $m$. 

This process is repeated for $g$ generations. With each iteration of the genetic algorithm, the population is better suited to the environment, {\it i.e.}, genes with high fitness scores are passed on to future generations. In each system, our genetic algorithm is employed 10 times, creating an ensemble of optimizations that effectively explore the entire search space. The individual that yielded the highest fitness score from the ensemble is selected as the optimum, and the parameter values corresponding to that individual represents the best fit.

In this work, an individual of the population is a $\mathtt{FREYA}$ parameter set encoded as a binary string representing a gene sequence, such as 
\begin{widetext}
\begin{center}
\begin{tabular}{ccccc}
$0100100111010$ & $100110001$ & $11101001101$ & $0000101101$ & $00101110010001$ \\
$e_0$ & $x$ & $c$ & $c_S$ & $d$TKE \\
\end{tabular}
\end{center}
\end{widetext}
We note that a non-uniform number of digits is assigned to each parameter in order to achieve a precision of three significant figures after the decimal place in the optimized values, giving a total number of features, the length of the gene sequence (binary string), $f = 58$.  The fitness score, $F$, of each individual is calculated as the inverse of our objective function, $F=1/\chi^2$. 

There are many ways the crossover operation can be applied to create the children for the next generation. In a one-point crossover, for example, the genes of two parents are swapped at a randomly chosen point in the gene sequence. Similarly, multi-point crossovers, where genes are swapped between pairs of randomly chosen points in the gene sequence, provide additional options. We have chosen to employ a uniform crossover operation, where a child is equally likely to inherit a gene from either parent at each point in its gene sequence. The mutation operation that follows, if applied, simply flips bits in the gene sequence of the children ($0 \leftrightarrow 1$). 

In order to prevent falling into local minima caused by a loss of diversity in the gene pool, we implemented a procedure to increase the mutation rate once the population diversity, quantified by the average Hamming distance, drops below a threshold $D_{\rm min}$. We found setting $D_{\rm min}=0.2f$ to be effective. Thus, the mutation rate is boosted if, on average, individuals in the population share 80\% or more of their genes. The rest of the hyperparameter values of the genetic algorithm were adopted from the guidelines set forth in Ref.~\cite{Goldberg1989}. We found that the following values yielded suitable results for our analysis: $p=50$, $k=3$, $c=0.80$, $m=1/f=1/58$, and $g=50$. To assess the sensitivity of our optimization results to the selected hyperparameters, we perturbed the hyperparameters around their nominal values for $^{252}$Cf(sf), the isotope with the most data available. The quality of the fit was found to be largely unaffected by these perturbations, demonstrating the robustness of the optimization with respect to the selected hyperparameters (see Appendix~\ref{sec:B}).

Genetic algorithms have the advantage of being parallelizable on a high performance computing platform. In the case of computationally heavy objective functions ({\it i.e.}, those with computational times of several minutes or longer), parallelizing the optimization can significantly reduce the optimization time. Because the genetic algorithm generates an entire population at once, it has the potential to calculate the fitness score of (or, equivalently, evaluate the objective function for) each individual simultaneously. Individuals in a population do not thus directly depend on one another. This is not true, for example, in the case of simulated annealing where each step in the random walk is dependent on the previous step, determined by evaluating the objective function~\cite{VanDyke2019}. Also, because the genetic algorithm can be run in parallel for the same compound nucleus, it is advantageous for future studies of the energy dependence of neutron-induced fission relative to simulated annealing which requires serial execution.

In order to optimize the parameters, an algorithm that samples the full parameter space evenly without being trapped in local minima must be employed. Since a population is randomly initialized within the parameter bounds to start the search, genetic algorithms explore the entire search space impartially. Therefore, genetic algorithms have a higher resistance to bias arising from an initial guess. Search algorithms utilizing a random walk require an initial starting point, which can introduce unconscious bias in the search, even if chosen randomly. Such inherently biased initial guesses can produce results far from the true minimum. Employing a genetic algorithm avoids using a biased starting point. 

\begin{table}
\caption{Ranges of parameter values assumed in the fits.}
\label{tab:ranges}
\begin{ruledtabular}
\begin{tabular}{cccccc}
$e_0$ (MeV$^{-1}$) & $x$ & $c$ & $c_S$ & $d$TKE (MeV) \\
\colrule
$[7, 12]$ & $[1.0, 1.5]$ & $[1, 3]$ & $[0.5, 1.5]$ & $[-5, 5]$ \\
\end{tabular}
\end{ruledtabular}
\end{table}

The parameter ranges used in Ref.~\cite{VanDyke2019} (shown in Table~\ref{tab:ranges}) have been adopted for our optimization. Although Ref.~\cite{VanDyke2019} only considered spontaneous fission, the same ranges have been used here for thermal neutron-induced fission. Each fissioning nucleus has been treated individually, with all five parameters varying independently regardless of the number of data sets available to constrain them. This is not unreasonable, as we do not expect the $\mathtt{FREYA}$ parameters to remain constant across nuclei. 

An ensemble of genetic algorithm optimizations is generated for each isotope and the resulting standard error on the optimized parameters is calculated from the ensemble. We report this quantity as the parameter uncertainty. The standard error reflects the stochastic nature inherent in a genetic algorithm. The parameter values reported in Sec.~\ref{sec:results} yielded the lowest $\chi^2$ from the ensemble for a given isotope. We note that we can reduce the uncertainties on our optimized parameters by generating further samples whereas the approach used in Ref.~\cite{VanDyke2019} cannot. 

The parameter uncertainties have been propagated through to the $\mathtt{FREYA}$ output. First, $\mathtt{FREYA}$ output is generated for each parameter set resulting from the ensemble obtained as described in the previous paragraph. The standard error for each bin of the $\mathtt{FREYA}$ observables is then calculated from the ensemble of $\mathtt{FREYA}$ output. This quantity is added to the statistical uncertainty in the $\mathtt{FREYA}$ calculation to yield the total uncertainty on the $\mathtt{FREYA}$ results shown in Sec.~\ref{sec:results}.

\section{Data employed in the fits}
\label{sec:data}

As noted in Sec.~\ref{sec:methods}, any fitting procedure relies on the objective function which computes how closely the $\mathtt{FREYA}$ output reproduces available experimental or evaluated data. We now discuss the sources and quality of the data employed in this study for spontaneous fission of $^{238}$U, $^{238,240,242}$Pu, $^{244}$Cm, and $^{252}$Cf and thermal neutron-induced fission of $^{233,235}$U and $^{239,241}$Pu.

\subsection{Spontaneous fission}

\begin{table}
\caption{Data employed in the spontaneous fission optimizations. Note that the evaluation of Ref.~\cite{Santi-Miller2008} provides the neutron multiplicity distribution, $P(\nu)$, as well as its first three moments: $\overline \nu$, $\nu_2$, and $\nu_3$. The measurement in Ref.~\cite{Chyzh2012} gives values for $\overline N_\gamma$, $\overline \epsilon_\gamma$, and $P(N_\gamma)$, whereas Ref.~\cite{Oberstedt2016} only gives $\overline N_\gamma$ and $\overline \epsilon_\gamma$.}
\label{tab:sfdata}
\begin{ruledtabular}
\begin{tabular}{lcccc}
Reaction & \# Data Sets & \# Evaluations & \# Observables \\
\colrule
$^{238}$U(sf) & - & 1~\cite{Santi-Miller2008} & 4 \\
\colrule
$^{238}$Pu(sf) & - & 1~\cite{Santi-Miller2008} & 4 \\
\colrule
$^{240}$Pu(sf) & 2~\cite{Oberstedt2016, Gerasimenko2002} & 1~\cite{Santi-Miller2008} & 7 \\
\colrule
$^{242}$Pu(sf) & 2~\cite{Oberstedt2016, Gerasimenko2002} & 1~\cite{Santi-Miller2008} & 7 \\
\colrule
$^{244}$Cm(sf) & 2~\cite{Schmidt1983, Boykov1997} & 1~\cite{Santi-Miller2008} & 6 \\
\colrule
$^{252}$Cf(sf) & 3~\cite{Chyzh2012, Dushin2004, Budtz-Jorgensen1988} & 2~\cite{Santi-Miller2008, Mannhart1987} & 10 \\
\end{tabular}
\end{ruledtabular}
\end{table}

Table~\ref{tab:sfdata} summarizes the data used in our fits for spontaneous fission.  The same data as employed in Ref.~\cite{VanDyke2019} are used here, with the addition of the PFNS data from Ref.~\cite{Gerasimenko2002} for $^{240,242}$Pu(sf). These data are presented as the ratio to the $^{252}$Cf(sf) PFNS evaluated by Mannhart~\cite{Mannhart1987}. The PFNS itself is obtained by multiplying the ratio data by the Mannhart evaluation.  Also note that while Ref.~\cite{VanDyke2019} inflated the uncertainties on $P(\nu)$, we have used the uncertainties reported in the Santi-Miller evaluation~\cite{Santi-Miller2008}, with the exception of $^{252}$Cf(sf).  This isotope, with many high-statistics measurements available, was used to benchmark our genetic algorithm against the simulated annealing approach of Ref.~\cite{VanDyke2019}.  To facilitate a direct comparison between the two methods, we therefore inflate the Santi-Miller uncertainties for $^{252}$Cf(sf) in the same manner as Ref.~\cite{VanDyke2019}. See Ref.~\cite{VanDyke2019} for a more thorough discussion of the other data used in the spontaneous fission fits.

While we have not included any other new data in these fits, aside from those of Ref.~\cite{Gerasimenko2002}, we did consider including $n$--$n$ correlation measurements for $^{252}$Cf(sf)~\cite{Gagarski2008} and $^{240}$Pu(sf)~\cite{Verbeke2018nn}. These data gated the angular correlations on several minimum outgoing neutron energies. We were particularly interested in employing such data to help constrain the $x$ parameter in the absence of $\overline \nu(A)$ data for $^{240}$Pu(sf).

As a first check, we tried adding the correlation data from Ref.~\cite{Gagarski2008} to the $^{252}$Cf(sf) fit.  We found that including $n$--$n$ correlation data here, where $\overline \nu(A)$ data are available to constrain $x$, did not change the fitted parameters within uncertainties.  Three cases were compared: the original result with $\overline \nu(A)$ data alone, as in Ref.~\cite{VanDyke2019}; a fit with $n$--$n$ correlation data from Ref.~\cite{Gagarski2008} alone; and a fit using both $\overline \nu(A)$ and $n$--$n$ correlations. Not only did the parameters not change significantly among these three cases, the quality of the comparison to the Mannhart PFNS was not degraded.  

However, when the correlation data from Ref.~\cite{Verbeke2018nn} were included in the $^{240}$Pu(sf) fit, these data were in considerable tension with the PFNS data, significantly reducing the quality of the agreement with the PFNS.  Such tension could arise from differences in the PFNS near the energy thresholds employed in the $n$--$n$ correlation measurement.  Because the PFNS is more important for applications, we excluded the $n$--$n$ correlation data from the final $^{240}$Pu(sf) fit. Nonetheless, comparisons to some of these data are provided in Sec.~\ref{sec:results}.  We could revisit including such correlations in future fits of this isotope if new PFNS measurements supersede those of Ref.~\cite{Gerasimenko2002} or more $n$--$n$ correlation data become available.

The yields and TKE$(A_H)$ for spontaneous fission are taken from data~\cite{Hambsch1997, Schmidt1985, EnglandRider1985, Schillebeeckx:1992, Ivanov1985}.  Because many of these yields lack high statistical accuracy, such as those for $^{238,240,242}$Pu(sf)~\cite{Schillebeeckx:1992}, we considered taking these quantities from GEF~\cite{gef2025}.  However, we found that, despite greater statistical precision, GEF gave zero yield at symmetric fission with narrower asymmetric mass yields than the data.  In addition, the calculated TKE$(A_H)$ had a significantly steeper slope for $A_H> 132$. Thus we chose to sample $Y(A)$ and TKE$(A_H)$ from the experimental data despite the lower statistics.

\subsection{Thermal neutron-induced fission}

\begin{table}
\caption{Data employed in fits for thermal neutron-induced fission of $^{233,235}$U and $^{239,241}$Pu. Note that the evaluation of Ref.~\cite{Holden-Zucker1988} provides the neutron multiplicity distribution, $P(\nu)$, as well as its first three moments: $\overline \nu$, $\nu_2$, and $\nu_3$. Reference~\cite{Akindele2019}, on the other hand, gave the moments $\overline \nu$, $\nu_2$, and $\nu_3$ alone. References~\cite{Pleasonton1973} and ~\cite{Pleasonton1972} provide data on $\overline N_\gamma$, $\overline \epsilon_\gamma$, $\overline N_\gamma(A)$, $\overline \epsilon_\gamma(A)$, $\overline E_\gamma(A)$, and $\overline E_\gamma(\rm TKE)$, whereas Ref.~\cite{Chyzh2014} only provides data on $\overline N_\gamma$ and $\overline \epsilon_\gamma$.}
\label{tab:nfdata}
\begin{ruledtabular}
\begin{tabular}{lcccc}
Reaction & \# Data Sets & \# Evaluations & \# Observables \\
\colrule
$^{233}$U($n_{\rm th}$,f) & 3~\cite{Vorobyev2016, Pleasonton1973, Nishio1998:233U} & 1~\cite{Holden-Zucker1988} & 8 \\
\colrule
$^{235}$U($n_{\rm th}$,f) & 4~\cite{Pleasonton1972, Oberstedt2013, Nishio1998:235U, Vorobyev2009:235U} & 1~\cite{Holden-Zucker1988} & 12 \\
\colrule
$^{239}$Pu($n_{\rm th}$,f) & 2~\cite{Tsuchiya2000, Pleasonton1973} & 2~\cite{Holden-Zucker1988, Brown2018} & 10 \\
\colrule
$^{241}$Pu($n_{\rm th}$,f) & 2~\cite{Akindele2019, Chyzh2014} & - & 5 \\
\end{tabular}
\end{ruledtabular}
\end{table}

Table~\ref{tab:nfdata} summarizes the data used in our fits for thermal neutron-induced fission.  

The neutron multiplicity distribution, $P(\nu)$, as well as its first three moments for $^{233,235}$U($n_{\rm th}$,f) and $^{239}$Pu($n_{\rm th}$,f) are taken from the Holden-Zucker evaluation~\cite{Holden-Zucker1988}. While this evaluation is from the 1980s, it includes uncertainties whereas the more recent ENDF/B-VIII.0 evaluation~\cite{Brown2018} of $^{239}$Pu($n_{\rm th}$,f) does not. Aside from this, the differences between the two evaluations are minimal.  The $^{241}$Pu($n_{\rm th}$,f) data on $\overline\nu$, $\nu_2$, and $\nu_3$ are taken from Ref.~\cite{Akindele2019}. These neutron multiplicity data strongly influence the $c$ parameter.

The photon observables $\overline N_\gamma$, $\overline \epsilon_\gamma$, $\overline N_\gamma(A)$, $\overline \epsilon_\gamma(A)$, $\overline E_\gamma(A)$, and $\overline E_\gamma(\rm TKE)$ for $^{233}$U($n_{\rm th}$,f)  and $^{239}$Pu($n_{\rm th}$,f) were reported in Ref.~\cite{Pleasonton1973} while those for $^{235}$U($n_{\rm th}$,f) were presented in Ref.~\cite{Pleasonton1972}. These data are of varying statistical quality.  The measured $\overline N_\gamma(A)$, $\overline E_\gamma(A)$, $\overline \epsilon_\gamma (A)$ and $\overline E_\gamma({\rm TKE})$ generally have relatively large uncertainties and thus do not strongly influence the fit results.  These data were, nonetheless, included in the fits. The small measured uncertainties for $\overline E_\gamma$(TKE) from $^{233}$U($n_{\rm th}$,f) exceptionally dominate the fit, see Eq.~(\ref{eq:chisq}), and were not included to reduce this bias. Comparisons to all these data are shown in Appendix~\ref{sec:A}. These photon observables directly affect the parameter $c_S$.

In addition to the aforementioned photon data, the PFGS is also available for $^{235}$U($n_{\rm th}$,f)~\cite{Oberstedt2013} and $^{241}$Pu($n_{\rm th}$,f)~\cite{Chyzh2014}.  Reference~\cite{Chyzh2014} also reported the photon multiplicity distribution, $P(N_\gamma)$, for $^{241}$Pu($n_{\rm th}$,f). We found that including these data in the $^{241}$Pu($n_{\rm th}$,f) fit significantly degraded the quality of the fit to the moments of $P(\nu)$ from Ref.~\cite{Akindele2019}.  Thus the $P(N_\gamma)$ and PFGS data are not included in the fit for $^{241}$Pu($n_{\rm th}$,f).  We do, however, present a comparison to these data in Appendix~\ref{sec:A}. Instead, we extract the averages $\overline N_\gamma$ and $\overline \epsilon_\gamma$ from the $P(N_\gamma)$ and PFGS data, respectively, and include them in our fit. 

Experimental $\overline \nu(A)$ data are available for $^{233}$U($n_{\rm th}$,f)~\cite{Nishio1998:233U}, $^{235}$U($n_{\rm th}$,f)~\cite{Nishio1998:235U}, and $^{239}$Pu($n_{\rm th}$,f)~\cite{Tsuchiya2000}. References~\cite{Nishio1995:239Pu,Apalin1965} also measure the $\overline \nu(A)$ of $^{239}$Pu($n_{\rm th}$,f), but the uncertainties on these data are either absent or incomplete. Thus these $\overline \nu(A)$ data are not used to constrain the $x$ parameter. 

The $^{239}$Pu($n_{\rm th}$,f) PFNS, taken from the ENDF/B-VIII.0 evaluation~\cite{Brown2018}, covers outgoing neutron energies from 100~eV to 20~MeV. The PFNS uncertainties were assigned based on the mean uncertainties reported in Table~XX of the evaluation.  PFNS measurements for $^{233}$U($n_{\rm th}$,f) and $^{235}$U($n_{\rm th}$,f) are taken from Ref.~\cite{Vorobyev2016} and Ref.~\cite{Vorobyev2009:235U}, respectively.  All the $\mathtt{FREYA}$ parameters have some influence on the PFNS.

\section{Results}
\label{sec:results}

This section is divided into three parts.  The first presents the $\mathtt{FREYA}$ parameter values determined from our genetic algorithm for spontaneous and thermal neutron-induced fission and discusses potential applications to the Bohr hypothesis.  The second compares the $\mathtt{FREYA}$ output using our optimized parameters to the data employed in our fits.  The final part presents comparisons of the $\mathtt{FREYA}$ output using our optimized parameters to data that were not included in our fits.  In addition to comparisons with data, we also compare our $\mathtt{FREYA}$ results to those using the parameters from Ref.~\cite{VanDyke2019} for spontaneous fission and Ref.~\cite{Verbeke2018} for thermal neutron-induced fission. The corresponding $\chi_O^2$ is also given for each calculation. 

\subsection{Optimized Parameter Values}

Table ~\ref{tab:sfparameters} shows the spontaneous fission fit results using our genetic algorithm compared to the results from Ref.~\cite{VanDyke2019}. In the case of $^{252}$Cf(sf), where many high statistics measurements are available, the parameters agree within uncertainties between the two methods. We have therefore verified our genetic algorithm by confirming the results of the simulated annealing approach for $^{252}$Cf(sf). 

The parameter values obtained using our genetic algorithm for the remaining spontaneous fission cases vary more significantly from the results of Ref.~\cite{VanDyke2019}. Aside from $^{240}$Pu(sf) and $^{242}$Pu(sf), the same spontaneous fission data as Ref.~\cite{VanDyke2019} was used in our fits. We have improved the quality of the fits from Ref.~\cite{VanDyke2019}, as quantified by the smaller total $\chi^2$ for each isotope in Table~\ref{tab:sfparameters}. As discussed in Sec.~\ref{sec:methods}, genetic algorithms are less prone to local minima in the complex five-dimensional search space and do not require an initial guess that may bias the search.

The discrepancies between our fit results and those of Ref.~\cite{VanDyke2019} are somewhat expected for $^{240}$Pu(sf) and $^{242}$Pu(sf) as we have used PFNS data~\cite{Gerasimenko2002} in our fits that was not included in the fits of Ref.~\cite{VanDyke2019}. While some $\mathtt{FREYA}$ parameters have a direct correspondence to a particular observable, all the $\mathtt{FREYA}$ parameters have some effect on the PFNS, as we discuss now.

An increase in $e_0$ results in a larger scission temperature, which can lead to either increased neutron emission or emission of higher energy neutrons. In the latter case, the PFNS becomes ``harder,'' shifting emission towards higher outgoing neutron energies. Similarly, increasing $x$ shifts more of the statistical excitation energy at scission to the light fragment, which can again result in increased emission of higher energy neutrons. The parameter $c$ determines the width of the neutron multiplicity distribution. A narrower $P(\nu)$ could lead to fewer higher energy prompt neutrons emitted whereas a wider distribution could be the result of the emission of many low energy prompt neutrons. The $c_S$ parameter sets the degree of fluctuations around the mean value of angular momenta of the two fragments. Increasing $c_S$ shifts more of the excitation energy at scission to the rotational energy of the fragments, reducing the statistical excitation energy available for neutron emission. Finally, increasing $d$TKE corresponds to a decrease in the excitation energy available at scission, which can decrease the average neutron multiplicity while increasing the mean neutron energy. 

Including the PFNS for $^{240,242}$Pu(sf) may explain some shifts in the parameters obtained for these isotopes, as described in the previous paragraph, but cannot account for the parameter shifts relative to Ref.~\cite{VanDyke2019} for the other isotopes where no additional data are included. Parameter shifts may occur because we have used the reported uncertainties on the $P(\nu)$ data from Ref.~\cite{Santi-Miller2008} whereas the uncertainties on these data were inflated in Ref.~\cite{VanDyke2019}. Other shifts may occur for some isotopes because no data are available to more effectively constrain them, such as a lack of photon emission data to constrain $c_S$ or the absence of $\overline \nu(A)$ or $n$--$n$ correlation data to constrain $x$. 

\begin{figure}
    \includegraphics[width=1.0\linewidth]{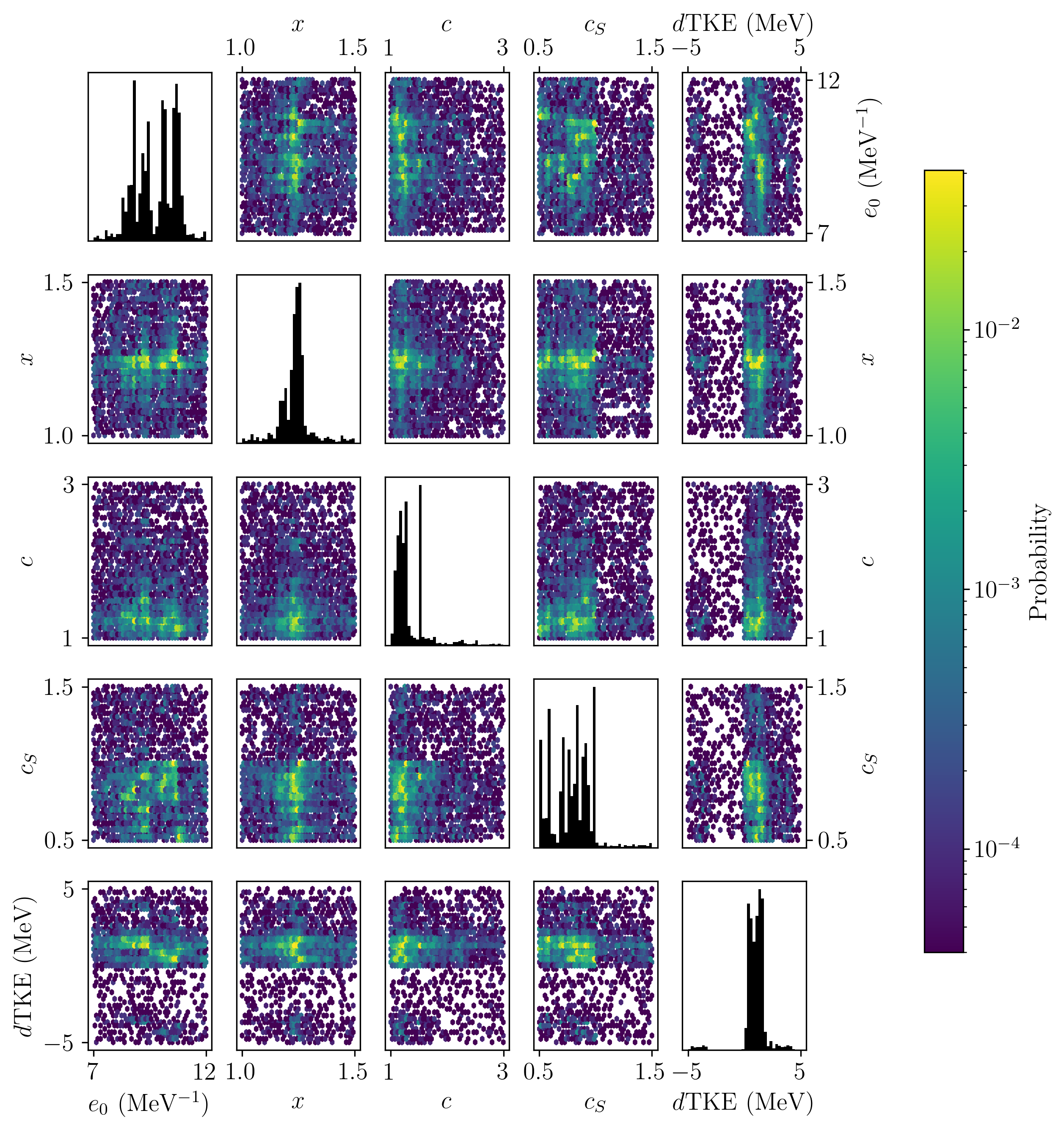}
    \caption{Probability distributions of the parameters sampled in the ensemble of $^{252}$Cf(sf) optimizations.}
    \label{fig:Cf252sf_param_dist}
\end{figure}

We note that $e_0$ and $c_S$ are the most difficult parameters to constrain because of their correlation.  We found that, for a given isotope, our ensemble of optimizations tended to encompass considerably wider ranges of values for $e_0$ and $c_S$ than for $x$, $c$, and $d$TKE (but still within the bounds set in Table~\ref{tab:ranges}). Figure~\ref{fig:Cf252sf_param_dist} illustrates this phenomenon for $^{252}$Cf(sf), the isotope with the most data available and therefore having the greatest number of parameter constraints.  The wider distributions of $e_0$ and $c_S$ reflect the fact that their strong correlation allows for more possible combinations of these two parameters, resulting in similar physics output.  The $e_0$ parameter sets the scission temperature, $T_{\rm sc}$, in Eq.~(\ref{eq:e0}) and $c_S$ is the ratio of the spin temperature, $T_S$, to $T_{\rm sc}$, as seen in Eq.~(\ref{eq:cS}).  Given  a fixed excitation energy at scission, a larger value of $e_0$ results in a larger $T_{\rm sc}$, which in turn can increase the probability of neutron emission.  To compensate, $c_S$ can increase to increase photon emission and keep the neutron emission fixed.  By the same token, a decrease in $c_S$ can be compensated by a decrease in $e_0$ in order to retain similar neutron emission.

While our $\chi^2$ for spontaneous fission is reduced relative to those in Ref.~\cite{VanDyke2019}, it can still be large unless the parameters are only fit to a single data set, as is the case for $^{238}$U(sf) and $^{238}$Pu(sf), both of which are constrained only by the $P(\nu)$ evaluation and its moments~\cite{Santi-Miller2008}.  When many observables are available for fitting, as is the case for $^{252}$Cf(sf) and thermal neutron-induced fission, except for $^{241}$Pu($n_{\rm th}$,f), the total $\chi^2$ can become large, even though fits to individual data sets are good, as seen in Eq.~(\ref{eq:total_chisq}). Furthermore, even small discrepancies between the $\mathtt{FREYA}$ output and data can result in large $\chi_O^2$ for high precision data, see Eq.~(\ref{eq:chisq}). Therefore, a large total $\chi^2$ for our results does not indicate a poor fit, as discussed in the subsequent subsections.

\begin{table*}
\centering
\caption{The optimized $\mathtt{FREYA}$ parameters and their corresponding $\chi^2$ for all spontaneously fissioning isotopes supported by $\mathtt{FREYA}$, along with the previous results from Ref.~\cite{VanDyke2019}. Because Ref.~\cite{VanDyke2019} does not report the $\chi^2$ for each reaction, we have calculated it using Eqs.~(\ref{eq:chisq}) and~(\ref{eq:total_chisq}) and included it here.}
\label{tab:sfparameters}
\begin{ruledtabular}
\begin{tabular}{lcccccc}
& $e_0$ (MeV$^{-1}$) & $x$ & $c$ & $c_S$ & $d$TKE (MeV) & $\chi^2$ \\

\colrule
\multicolumn{6}{l}{$^{238}$U(sf)} \\
\colrule

This work & $9.675 \pm 0.384$ & $1.338 \pm 0.063$ & $1.011 \pm 0.007$ & $1.029 \pm 0.055$ & $-1.512 \pm 0.102$ & $10.82 \pm 6.74$ \\
Ref.~\cite{VanDyke2019} & $10.391 \pm 0.352$ & $1.220 \pm 0.071$ & $0.929 \pm 0.283$ & $0.899 \pm 0.280$ & $-1.375 \pm 0.727$ & $16.31 \pm 8.09$ \\                 
\colrule
\multicolumn{6}{l}{$^{238}$Pu(sf)} \\
\colrule

This work & $7.318 \pm 0.492$ & $1.000 \pm 0.029$ & $1.922 \pm 0.028$ & $1.198 \pm 0.082$ & $-1.462 \pm 0.122$ & $5.04 \pm 4.64$ \\
Ref.~\cite{VanDyke2019} & $10.521 \pm 0.581$ & $1.232 \pm 0.221$ & $1.968 \pm 0.071$ & $0.893 \pm 0.071$ & $-1.408 \pm 3.424$ & $7.87 \pm 5.62$ \\

\colrule
\multicolumn{6}{l}{$^{240}$Pu(sf)} \\
\colrule

This work & $11.304 \pm 1.208$ & $1.267 \pm 0.139$ & $1.708 \pm 0.136$ & $1.469 \pm 0.136$ & $-3.833 \pm 0.401$ & $33.53 \pm 14.06$ \\
Ref.~\cite{VanDyke2019} & $10.750 \pm 0.138$ & $1.307 \pm 0.071$ & $3.176 \pm 0.355$ & $0.908 \pm 0.023$ & $-3.219 \pm 0.112$ & $761.83 \pm 52.32$ \\ 

\colrule
\multicolumn{6}{l}{$^{242}$Pu(sf)} \\
\colrule

This work & $7.092 \pm 0.573$ & $1.499 \pm 0.041$ & $2.093 \pm 0.044$ & $0.974 \pm 0.019$ & $-1.557 \pm 0.109$ & $196.55 \pm 82.92$ \\
Ref.~\cite{VanDyke2019} & $10.018 \pm 1.768$ & $1.144 \pm 0.152$ & $3.422 \pm 0.341$ & $0.911 \pm 0.257$ & $-1.662 \pm 0.118$ & $254.01 \pm 26.22$ \\

\colrule
\multicolumn{6}{l}{$^{244}$Cm(sf)} \\
\colrule

This work & $8.362 \pm 0.389$ & $1.478 \pm 0.012$ & $1.292 \pm 0.038$ & $0.714 \pm 0.079$ & $-3.954 \pm 0.121$ & $161.43 \pm 17.59$ \\
Ref.~\cite{VanDyke2019} & $10.488 \pm 1.519$ & $1.239 \pm 0.148$ & $1.291 \pm 0.582$ & $0.906 \pm 0.322$ & $-4.494 \pm 0.167$ & $339.77 \pm 9.82$ \\

\colrule
\multicolumn{6}{l}{$^{252}$Cf(sf)} \\
\colrule

This work & $10.430 \pm 0.496$ & $1.260 \pm 0.006$ & $1.185 \pm 0.021$ & $0.890 \pm 0.064$ & $0.508 \pm 0.293$ & $483.05 \pm 24.63$ \\
Ref.~\cite{VanDyke2019} & $10.429 \pm 1.090$ & $1.274 \pm 0.187$ & $1.191 \pm 0.362$ & $0.875 \pm 0.020$ & $0.525 \pm 0.078$ & $496.22 \pm 7.45$ \\

\end{tabular}
\end{ruledtabular}
\end{table*}

Table~\ref{tab:nfparameters} presents the optimized parameter values and uncertainties for thermal neutron-induced fission of $^{233,235}$U and $^{239,241}$Pu. We note that, prior to this work, universal values were largely assumed for most of the five parameters, with only $c$ and $d$TKE adjusted to fit $P(\nu)$ and $\overline \nu$, respectively, neglecting any isotope dependence. This work is the first attempt to capture the isotope dependence of the $\mathtt{FREYA}$ parameters for thermal neutron-induced fission.

The default $\mathtt{FREYA}$ parameters assumed in Ref.~\cite{Verbeke2018} are also shown in Table~\ref{tab:nfparameters}. None of the cases studied finds agreement with these default values within uncertainties.   The $c_S$ and $d$TKE parameters display the most significant discrepancies, whereas the values of $e_0$, $x$, and $c$ are generally closer to the assumed values. The $\chi^2$ has been calculated using Eqs.~(\ref{eq:chisq}) and~(\ref{eq:total_chisq}) for all parameter sets.  We have obtained significantly smaller $\chi^2$ by 1--2 orders of magnitude for all thermal neutron-induced fission relative to Ref.~\cite{Verbeke2018}.  This work has therefore improved the descriptive capabilities of $\mathtt{FREYA}$ for thermal neutron-induced fission.

\begin{table*}
\caption{The optimized $\mathtt{FREYA}$ parameters and their corresponding $\chi^2$ for all thermal neutron-induced fissioning isotopes supported by $\mathtt{FREYA}$, along with the default values from Ref.~\cite{Verbeke2018}. The $\chi^2$ for Ref.~\cite{Verbeke2018} has been calculated using Eqs.~(\ref{eq:chisq}) and~(\ref{eq:total_chisq}) and included here.}
\label{tab:nfparameters}
\begin{ruledtabular}
\begin{tabular}{lcccccc}
& $e_0$ (MeV$^{-1}$) & $x$ & $c$ & $c_S$ & $d$TKE (MeV) & $\chi^2$ \\

\colrule
\multicolumn{6}{l}{$^{233}$U($n_{\rm th}$,f)} \\
\colrule

This work & $11.243 \pm 0.342$ & $1.204 \pm  0.005$ & $1.250 \pm 0.013$ & $0.743 \pm 0.052$ & $-2.394 \pm 0.152$ & $101.13 \pm 7.04$ \\
Ref.~\cite{Verbeke2018} & $10.37$ & $1.15$ & $1.2$ & $0.87$ & $0.39$ & $2,975.08 \pm 96.07$ \\

\colrule
\multicolumn{6}{l}{$^{235}$U($n_{\rm th}$,f)} \\
\colrule

This work & $10.227 \pm 0.097$ & $1.176 \pm 0.010$ & $1.278 \pm 0.008$ & $1.469 \pm 0.009$ & $-2.442 \pm 0.022$ & $331.29 \pm 200.90$ \\
Ref.~\cite{Verbeke2018} & $10.37$ & $1.15$ & $1.3$ & $0.87$ & $0.39$ & $5,866.91 \pm 144.56$ \\

\colrule
\multicolumn{6}{l}{$^{239}$Pu($n_{\rm th}$,f)} \\
\colrule

This work & $10.287 \pm 0.315$ & $1.058 \pm 0.115$ & $1.153 \pm 0.039$ & $0.810 \pm 0.132$ & $-0.858 \pm 0.349$ & $611.20 \pm 494.27$ \\
Ref.~\cite{Verbeke2018} & $10.37$ & $1.15$ & $1.2$ & $0.87$ & $1.1$ & $30,533.24 \pm 642.15$ \\

\colrule
\multicolumn{6}{l}{$^{241}$Pu($n_{\rm th}$,f)} \\
\colrule

This work & $7.020 \pm 0.671$ & $1.225 \pm 0.034$ & $1.125 \pm 0.020$ & $1.130 \pm 0.012$ & $-1.457 \pm 0.219$ & $3.26 \pm 3.66$ \\
Ref.~\cite{Verbeke2018} & $10.37$ & $1.15$ & $1.2$ & $0.87$ & $1.1$ & $357.51 \pm 37.88$\\

\end{tabular}
\end{ruledtabular}
\end{table*}

\subsubsection{Application to the Bohr hypothesis}

The Bohr hypothesis~\cite{Bohr:1936zz} states that the decay of a compound nucleus for a given excitation energy, spin, and parity is independent of its formation.  Based on this hypothesis, Bohr and Wheeler~\cite{Bohr:1939ej} formulated the theory of nuclear fission employing the liquid drop model of the nucleus.  The theory predicts that the timescale for compound nucleus formation, $10^{-20}$--$10^{-15}$~s, is sufficiently long to allow the final-state nucleus to ``lose" the memory of how it was formed.  Thus a nucleus will fission according to its initial state regardless of how that initial state was prepared. 

The Bohr hypothesis has been tested at the level of cumulative fission yields at the Triangle Universities Nuclear Laboratory (TUNL) with a quasimonoenergetic neutron beam, $E_n = 4.6$~MeV, on $^{239}$Pu and an 11.2~MeV photon beam on $^{240}$Pu, resulting in the same compound nuclear final state, $^{240}$Pu, at a similar excitation energy~\cite{Tonchev:2017}.  The yields should agree if the Bohr hypothesis is correct, modulo uncertainties in the final state spin and parity of the compound nucleus in the two cases which could not be controlled for.  Most of the measured cumulative yields, obtained by gamma spectroscopy, were in good agreement.  Because the prompt independent yields cannot be measured by gamma spectroscopy, the TKE distribution was not obtained as a secondary check.

Further checks involving the decays of the fission fragments have not been performed.  Such checks are important since this assumption is employed in $\mathtt{FREYA}$ to calculate fission observables for photofission~\cite{Mueller:2014gxa,Clarke:2017hqv} and deuteron-induced fission~\cite{Gjestvang:2021isl}.

It is tempting to assume that the fragment pairs studied here: [$^{239}$Pu($n_{\rm th}$,f), $^{240}$Pu(sf)] and 
[$^{241}$Pu($n_{\rm th}$,f), $^{242}$Pu(sf)] could serve as a check of the Bohr hypothesis. However, while the compound nucleus is the same for these pairs, the excitation energy is not.  Spontaneous fission involves tunneling through the fission barrier, resulting in the fission of a ``cold" nucleus.  On the other hand, even thermal neutron-induced fission results in a finite initial excitation energy.  Thus the yields and TKE$(A_H)$ of the partner nuclei in each pair are not in agreement with each other.  The resulting neutron and photon observables are also not the same. For example, Akindele {\it et al.}~\cite{Akindele2019} measured the neutron multiplicity moments of $^{241}$Pu($n_{\rm th}$,f), finding $\overline \nu = 2.93$, $\nu_2 = 6.60$, and $\nu_3 = 12.50$.  On the other hand, the Santi-Miller evaluation of  $^{242}$Pu(sf) gives much smaller values, $\overline \nu = 2.149$, $\nu_2 = 3.809$, and $\nu_3 = 5.349$~\cite{Santi-Miller2008}.  

However, one would expect that, if the Bohr hypothesis were correct, measurements like those of Ref.~\cite{Tonchev:2017} comparing photofission and neutron-induced fission at incident energies that produce the same compound nucleus at similar initial excitation energies should produce the same TKE and result in comparable neutron and photon multiplicities and spectra.  Of course, while one can control the choice of compound nucleus as well as, at least to some degree, the initial excitation energy by controlling or measuring the incident neutron and photon energies, it is not possible to fix the spin and parity of the compound nucleus prior to scission.  To check this correspondence between final-state observables, one would need to measure the fission fragment yields as well as prompt neutron and photon emission, requiring a significantly different experimental setup than that of Tonchev {\it et al.}~\cite{Tonchev:2017}.  Validating the Bohr hypothesis would validate the working assumption of $\mathtt{FREYA}$ for other systems and enhance its predictive capabilities.

\subsection{Comparison to calibration data}

The optimized parameters presented in Tables~\ref{tab:sfparameters} and~\ref{tab:nfparameters} are now used to generate sets of one million $\mathtt{FREYA}$ events and the results are compared to data included in our fits. Direct comparisons are presented as well as ratios of the calculated to experimental or evaluated values, C/E. 

The uncertainties on the $\mathtt{FREYA}$ results using our new fits reflect both the statistical uncertainty in the $\mathtt{FREYA}$ calculation as well as the propagation of the uncertainties on the parameters as described in Sec.~\ref{sec:methods}. Aside from quantities with very low statistics, such as the PFNS above $\sim10$~MeV, the parameter uncertainties dominate the calculated uncertainties.  The maximum contribution of the statistical uncertainty to the total uncertainty was found to be less than $5\%$.

We also show the $\mathtt{FREYA}$ results using the parameter values determined in Ref.~\cite{VanDyke2019} for spontaneous fission or those given in Ref.~\cite{Verbeke2018} for thermal neutron-induced fission, as well as the corresponding $\chi_O^2$. The uncertainties on these results include only the statistical uncertainty in the $\mathtt{FREYA}$ calculation.  Thus the uncertainties on the spontaneous fission results, aside from the average values in Tables~\ref{tab:sf-neutron-comparison} and \ref{tab:sf-photon-comparison} which are taken directly from Ref.~\cite{VanDyke2019}, typically significantly underestimate the actual uncertainties.  On the other hand, no uncertainties were given for the parameters in Ref.~\cite{Verbeke2018}.

\subsubsection{Spontaneous fission}

\begin{table*}
\caption{Average neutron multiplicity, $\overline \nu$, as well as the second and third moments of the neutron multiplicity distribution, $\nu_2$ and $\nu_3$, for spontaneous fission compared with the data employed in the fits. The values from Ref.~\cite{VanDyke2019} are also shown.}
\label{tab:sf-neutron-comparison}
\begin{ruledtabular}
\begin{tabular}{llccccc} 
 & & & \multicolumn{2}{c}{This work} & \multicolumn{2}{c}{Ref.~\cite{VanDyke2019}} \\
 
Reaction & $\nu_n$ & Literature & $\mathtt{FREYA}$ & C/E & $\mathtt{FREYA}$ & C/E \\
\colrule

\colrule
\multirow{3}{*}{$^{238}$U(sf)}

 & $\overline \nu$ & $1.98 \pm 0.03$ & $1.979 \pm 0.002$ & $1.00 \pm 0.02$ & $2.00 \pm 0.94$ & $1.01 \pm 0.22$ \\
 & $\nu_2$ & $2.874 \pm 0.141$ & $2.867 \pm 0.009$ & $1.00 \pm 0.05$ & $2.87 \pm 3.37$ & $1.00 \pm 1.27$ \\
 & $\nu_3$ & $2.822 \pm 0.481$ & $2.950 \pm 0.015$ & $1.05 \pm 0.18$ & $2.83 \pm 9.81$ & $1.00 \pm 11.71$ \\
       
\colrule
\multirow{3}{*}{$^{238}$Pu(sf)}

 & $\overline \nu$ & $2.19 \pm 0.07$ & $2.187 \pm 0.008$ & $1.00 \pm 0.03$ & $2.17 \pm 1.15$ & $0.99 \pm 0.27$ \\
 & $\nu_2$ & $3.874 \pm 0.194$ & $3.903 \pm 0.008$ & $1.01 \pm 0.05$ & $3.85 \pm 4.45$ & $0.99 \pm 1.26$ \\
 & $\nu_3$ & $5.417 \pm 0.271$ & $5.396 \pm 0.016$ & $1.00 \pm 0.05$ & $5.25 \pm 10.97$ & $0.97 \pm 4.10$ \\

\colrule
\multirow{3}{*}{$^{240}$Pu(sf)}

 & $\overline \nu$ & $2.154 \pm 0.005$ & $2.154 \pm 0.004$ & $1.00 \pm 0.01$ & $2.22 \pm 1.25$ & $1.03 \pm 0.33$ \\
 & $\nu_2$ & $3.789 \pm 0.029$ & $3.794 \pm 0.018$ & $1.01 \pm 0.01$ & $4.26 \pm 4.88$ & $1.12 \pm 1.66$ \\
 & $\nu_3$ & $5.211 \pm 0.149$ & $5.150 \pm 0.070$ & $0.99 \pm 0.03$ & $6.53 \pm 13.30$ & $1.25 \pm 6.51$ \\

\colrule
\multirow{3}{*}{$^{242}$Pu(sf)}

 & $\overline \nu$ & $2.149 \pm 0.008$ & $2.148 \pm 0.006$ & $1.00 \pm 0.01$ & $2.12 \pm 1.19$ & $0.99 \pm 0.30$ \\ 
 & $\nu_2$ & $3.809 \pm 0.036$ & $3.856 \pm 0.014$ & $1.01 \pm 0.01$ & $3.79 \pm 4.51$ & $0.99 \pm 1.40$ \\
 & $\nu_3$ & $5.349 \pm 0.036$ & $5.347 \pm 0.033$ & $1.00 \pm 0.01$ & $5.36 \pm 12.13$ & $1.00 \pm 5.14$ \\

\colrule
\multirow{3}{*}{$^{244}$Cm(sf)}

 & $\overline \nu$ & $2.71 \pm 0.01$ & $2.716 \pm 0.001$ & $1.00 \pm 0.01$ & $2.70 \pm 1.16$ & $1.00 \pm 0.18$ \\
 & $\nu_2$ & $5.941 \pm 0.019$ & $5.945 \pm 0.005$ & $1.00 \pm 0.01$ & $5.95 \pm 5.46$ & $1.00 \pm 0.84$ \\
 & $\nu_3$ & $10.112 \pm 0.175$ & $9.957 \pm 0.002$ & $0.99 \pm 0.02$ & $10.17 \pm 16.78$ & $1.01 \pm 2.75$ \\

\colrule
\multirow{3}{*}{$^{252}$Cf(sf)}

 & $\overline \nu$ & $3.757 \pm 0.010$ & $3.763 \pm 0.024$ & $1.00 \pm 0.01$ & $3.74 \pm 1.30$ & $1.00 \pm 0.13$ \\
 & $\nu_2$ & $11.952 \pm 0.019$ & $12.125 \pm 0.138$ & $1.01 \pm 0.01$ & $11.94 \pm 8.79$ & $1.00 \pm 0.54$ \\
 & $\nu_3$ & $31.668 \pm 0.175$ & $32.777 \pm 0.505$ & $1.04 \pm 0.02$ & $31.84 \pm 39.94$ & $1.01 \pm 1.59$ \\

\end{tabular}
\end{ruledtabular}
\end{table*}

\begin{table*}
\caption{Average photon multiplicity, $\overline N_\gamma$, and average photon energy, $\overline \epsilon_\gamma$, for spontaneous fission compared with the data employed in the fits. The values from Ref.~\cite{VanDyke2019} are also shown.}
\label{tab:sf-photon-comparison}
\begin{ruledtabular}
\begin{tabular}{llccccc} 
 & & & \multicolumn{2}{c}{This work} & \multicolumn{2}{c}{Ref.~\cite{VanDyke2019}} \\ 

Reaction & Average & Literature & $\mathtt{FREYA}$ & C/E & $\mathtt{FREYA}$ & C/E \\

\colrule
\multirow{2}{*}{$^{238}$U(sf)}

 & $\overline N_\gamma$ & $-$ & $6.796 \pm 0.431$ & $-$ & $6.49 \pm 2.42$ & $-$ \\
 & $\overline \epsilon_\gamma$ (MeV) & $-$ & $0.923 \pm 0.023$ & $-$ & $0.94 \pm 0.87$ & $-$ \\
       
\colrule
\multirow{2}{*}{$^{238}$Pu(sf)}

 & $\overline N_\gamma$ & $-$ & $7.147 \pm 0.533$ & $-$ & $6.47 \pm 2.43$ & $-$ \\
 & $\overline \epsilon_\gamma$ (MeV) & $-$ & $0.993 \pm 0.037$ & $-$ & $1.05 \pm 0.93$ & $-$ \\

\colrule
\multirow{2}{*}{$^{240}$Pu(sf)}

 & $\overline N_\gamma$ & $8.2 \pm 0.4$ & $8.026 \pm 0.324$ & $0.98 \pm 0.06$ & $6.60 \pm 2.48$ & $0.80 \pm 0.09$ \\
 & $\overline \epsilon_\gamma$ (MeV) & $0.80 \pm 0.07$ & $0.963 \pm 0.013$ & $1.20 \pm 0.11$ & $1.0 \pm 0.91$ & $1.24 \pm 1.26$ \\

\colrule
\multirow{2}{*}{$^{242}$Pu(sf)}

 & $\overline N_\gamma$ & $6.72 \pm 0.07$ & $6.824 \pm 0.140$ & $1.02 \pm 0.02$ & $6.61 \pm 2.43$ & $0.98 \pm 0.13$ \\
 & $\overline \epsilon_\gamma$ (MeV) & $0.843 \pm 0.012$ & $0.895 \pm 0.023$ & $1.06 \pm 0.03$ & $0.96 \pm 0.89$ & $1.14 \pm 1.12$ \\

\colrule
\multirow{2}{*}{$^{244}$Cm(sf)}
    
 & $\overline N_\gamma$ & $-$ & $6.616 \pm 0.002$ & $-$ & $7.07 \pm 2.56$ & $-$ \\
 & $\overline \epsilon_\gamma$ (MeV) & $-$ & $1.008 \pm 0.001$ & $-$ & $1.01 \pm 0.93$ & $-$ \\

\colrule
\multirow{2}{*}{$^{252}$Cf(sf)}

 & $\overline N_\gamma$ & $8.14 \pm 0.40$ & $7.792 \pm 0.571$ & $0.96 \pm 0.08$ & $7.71 \pm 2.80$ & $0.95 \pm 0.12$ \\
 & $\overline \epsilon_\gamma$ (MeV) & $0.94 \pm 0.05$ & $0.910 \pm 0.019$ & $0.97 \pm 0.06$ & $0.91 \pm 0.86$ & $0.97 \pm 0.83$ \\

\end{tabular}
\end{ruledtabular}
\end{table*}

Table~\ref{tab:sf-neutron-comparison} compares the first three moments of the neutron multiplicity distribution calculated from the $\mathtt{FREYA}$ output with the evaluated results for spontaneous fission. The parameter values determined in this work reproduce the average neutron multiplicity, $\overline \nu$, and the second and third moments of the neutron multiplicity distribution, $\nu_2$ and $\nu_3$, with high precision for all reactions.  The parameters from Ref.~\cite{VanDyke2019} also reproduce these data to within a few percent except for $^{240}$Pu(sf). 

In Table~\ref{tab:sf-photon-comparison}, the average photon multiplicity, $\overline N_\gamma$, and the average photon energy, $\overline \epsilon_\gamma$, calculated from the $\mathtt{FREYA}$ output are compared to measured data for spontaneous fission, where available.  While the $\mathtt{FREYA}$ results differ more significantly from these data than the $P(\nu)$ moments in Table~\ref{tab:sf-neutron-comparison}, the C/E ratio is near unity in most cases. The photon data tend to have larger relative uncertainties, giving these observables less ``weight'' in our optimization, see Eq.~(\ref{eq:chisq}). We note that the parameters determined in this work have yielded improved or equivalent agreement with all the spontaneous fission photon data in Table~\ref{tab:sf-photon-comparison} compared to those calculated with the parameters of Ref.~\cite{VanDyke2019}.

Taken together, Tables~\ref{tab:sf-neutron-comparison} and~\ref{tab:sf-photon-comparison} validate our genetic algorithm relative to the simulated annealing approach used in Ref.~\cite{VanDyke2019}.  Additionally, the uncertainties on our results are significantly reduced relative to those of Ref.~\cite{VanDyke2019}. 
 
\begin{figure}
    \includegraphics[width=1.0\linewidth]{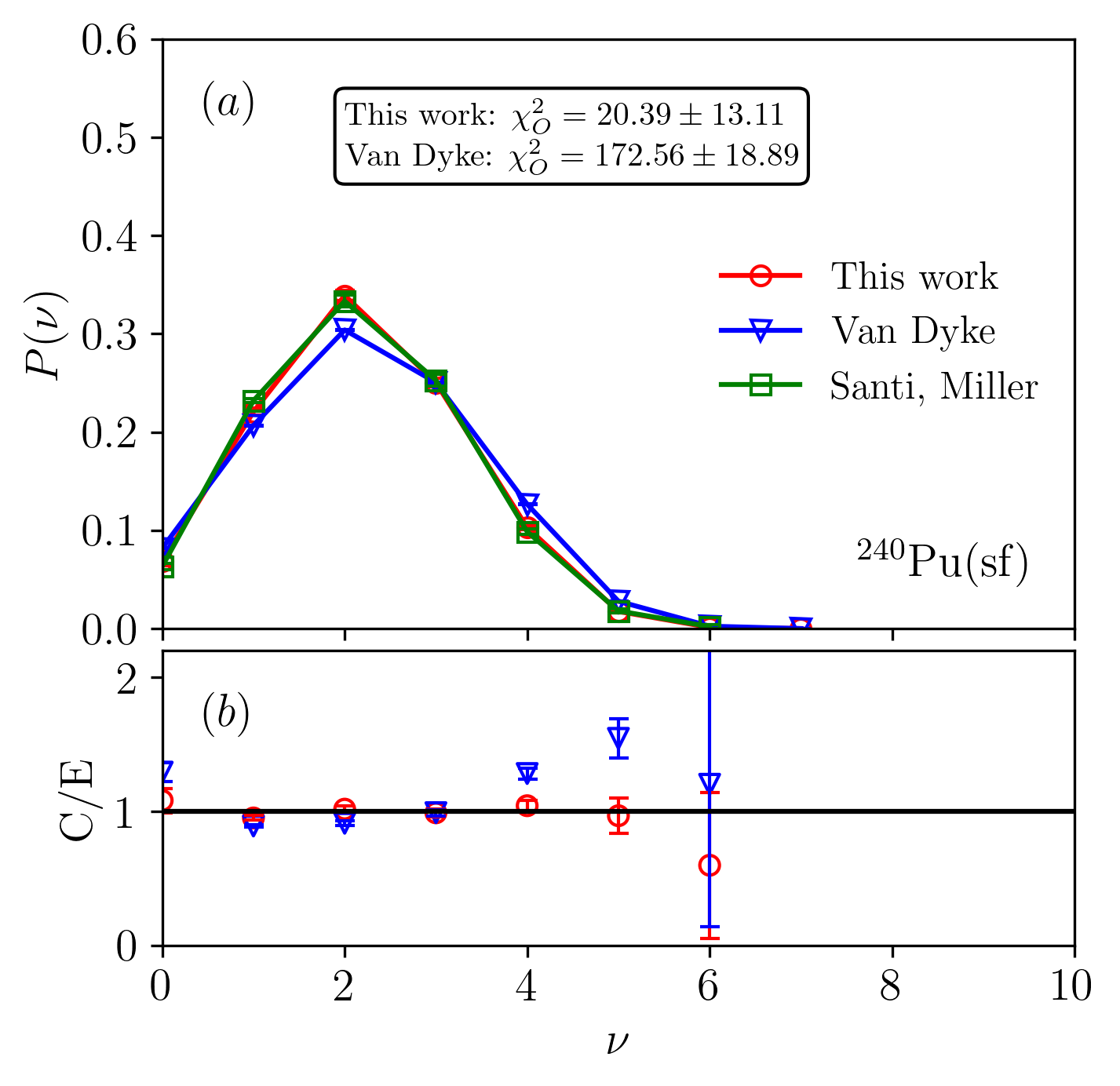}
    \caption{(a) The neutron multiplicity distribution for $^{240}$Pu(sf) using the parameter values from this work and those of Ref.~\cite{VanDyke2019} shown with the Santi-Miller evaluation~\cite{Santi-Miller2008}. (b) Ratio of $\mathtt{FREYA}$ calculations to the evaluation.}
    \label{fig:Pu240sf_P(nu)}
\end{figure}

As noted in Sec.~\ref{sec:data}, we have
used the reported uncertainties on the $P(\nu)$ data from Ref.~\cite{Santi-Miller2008}, whereas the uncertainties on these data were inflated in Ref.~\cite{VanDyke2019}. Figure~\ref{fig:Pu240sf_P(nu)} shows the neutron multiplicity distribution, $P(\nu)$, for $^{240}$Pu(sf) from $\mathtt{FREYA}$ compared to the Santi-Miller evaluation. The evaluated distribution is better reproduced using the parameters determined in this work compared to those from Ref.~\cite{VanDyke2019}, as is quantified by the order of magnitude smaller $\chi_O^2$. 

\begin{figure}
    \includegraphics[width=1.0\linewidth]{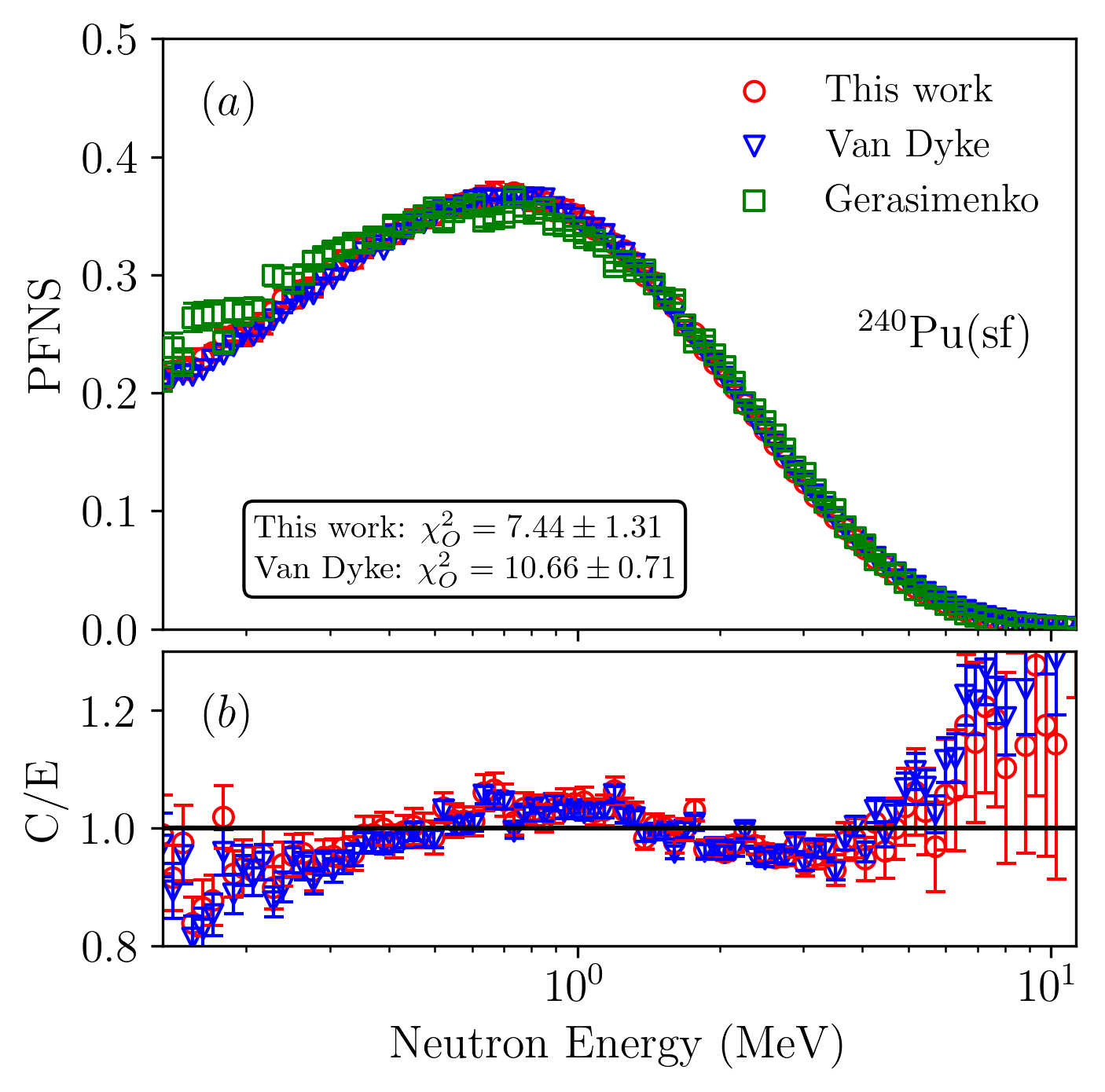}
    \caption{(a) Neutron energy spectrum for $^{240}$Pu(sf) using the parameter values from this work and  from Ref.~\cite{VanDyke2019} compared to experimental data from Ref.~\cite{Gerasimenko2002}. Note the logarithmic scale on the $x$-axis. (b) Ratio of $\mathtt{FREYA}$ calculations to the experimental data, C/E.}
    \label{fig:Pu240sf_PFNS}
\end{figure}

Furthermore, our $^{240}$Pu(sf) and $^{242}$Pu(sf) fits have been further constrained by the PFNS data from Ref.~\cite{Gerasimenko2002} relative to Ref.~\cite{VanDyke2019}. The PFNS is compared to the $^{240}$Pu(sf) data in Fig.~\ref{fig:Pu240sf_PFNS}.  The spectral data from Ref.~\cite{Gerasimenko2002} is not completely smooth, particularly for outgoing neutron energies below 1~MeV. It is therefore not surprising that the $\mathtt{FREYA}$ result does not exactly reproduce the data in this region. Within the uncertainties ${\rm C/E} \sim 1$ for nearly all outgoing neutron energies, for both the parameters determined in this work and those of Ref.~\cite{VanDyke2019}. However, the parameter values from this work have achieved a smaller $\chi_O^2$ than the previously determined values, with better agreement at the highest outgoing neutron energies in particular.   

\begin{figure}
    \includegraphics[width=1.0\linewidth]{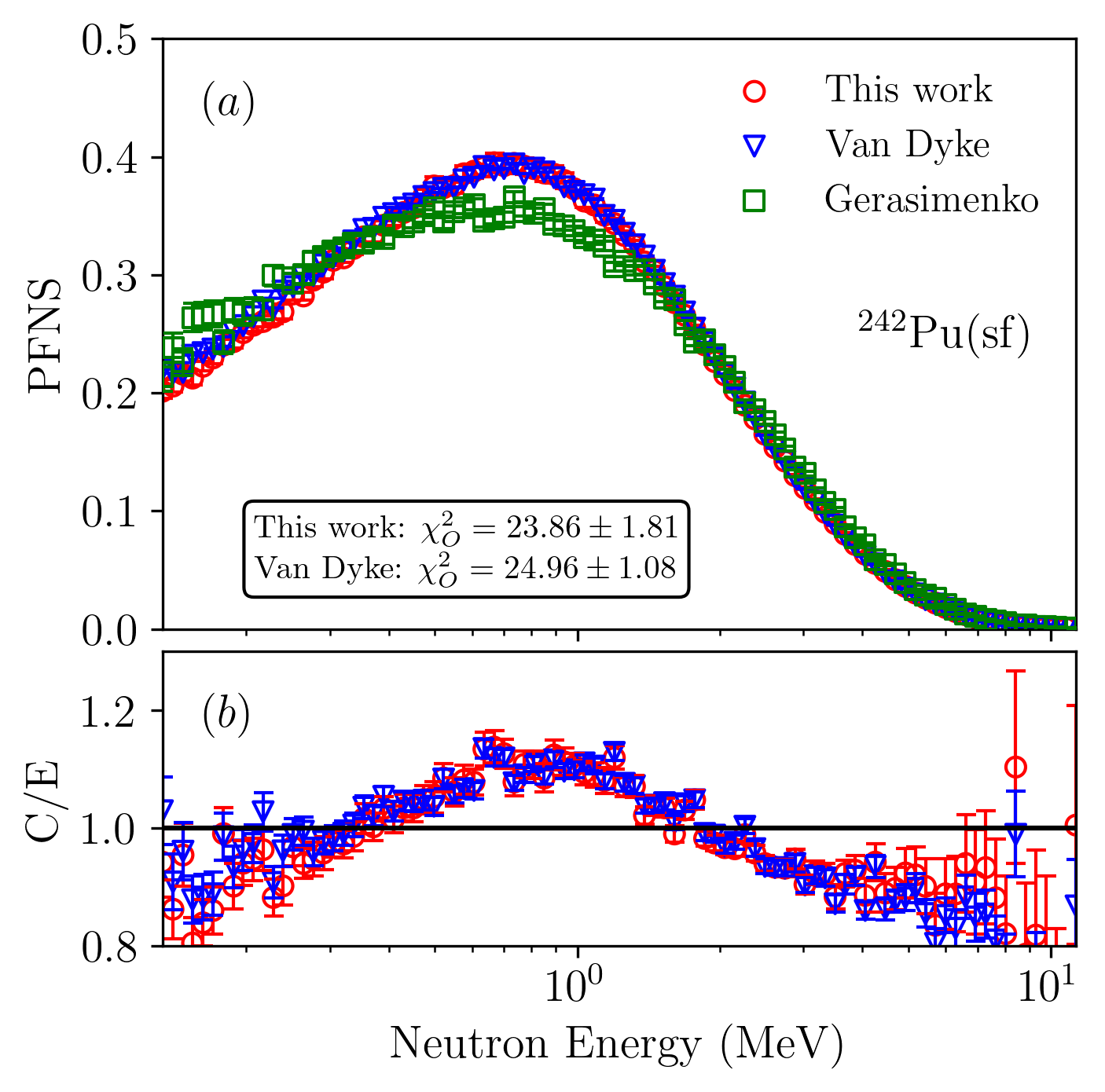}
    \caption{Same as Fig.~\ref{fig:Pu240sf_PFNS} but for $^{242}$Pu(sf).}
    \label{fig:Pu242sf_PFNS}
\end{figure}

In Fig.~\ref{fig:Pu242sf_PFNS} the PFNS calculated from the $\mathtt{FREYA}$ output is compared to the measured data for $^{242}$Pu(sf). At low outgoing neutron energies, near the ``peak'' of the PFNS, both calculations overshoot the data from Ref.~\cite{Gerasimenko2002}. Again, the data from Ref.~\cite{Gerasimenko2002} is not completely smooth at low outgoing neutron energies, making it difficult to reproduce the measured data in this region. At higher neutron energies the data are reproduced reasonably well up to $\sim7$~MeV. In this case, the parameter sets from this work and Ref.~\cite{VanDyke2019} yield $\chi_O^2$ that agree within uncertainties.


\subsubsection{Thermal neutron-induced fission}

We now turn to thermal neutron-induced fission of $^{233,235}$U and $^{239,241}$Pu, the primary new results of this work.  Table~\ref{tab:nf-neutron-comparison} compares the first three moments of the neutron multiplicity distribution calculated from the $\mathtt{FREYA}$ output with evaluated and experimental results, where applicable, for thermal neutron-induced fission. The parameters determined in this work reproduce the moments of the neutron multiplicity distribution rather well in all cases.  This is in stark contrast with the results using the assumed parameter values of Ref.~\cite{Verbeke2018}, which at best reproduces $\overline \nu$ within 7\%, becoming much worse for the higher moments of $P(\nu)$. 

\begin{table*}
\caption{Same as Table~\ref{tab:sf-neutron-comparison} but for thermal neutron-induced fission and the $\mathtt{FREYA}$ results using the default parameter values from Ref.~\cite{Verbeke2018}. Note that the uncertainties on the $\mathtt{FREYA}$ results from Ref.~\cite{Verbeke2018} are purely statistical, whereas we have, in addition to the statistical uncertainty, propagated the fit uncertainties through to our results.}
\label{tab:nf-neutron-comparison}
\begin{ruledtabular}
\begin{tabular}{llccccc} 
 & & & \multicolumn{2}{c}{This work} & \multicolumn{2}{c}{Ref.~\cite{Verbeke2018}} \\ 

Reaction & $\nu_n$ & Literature & $\mathtt{FREYA}$ & C/E & $\mathtt{FREYA}$ & C/E \\

\colrule
\multirow{3}{*}{$^{233}$U($n_{\rm th}$,f)}

 & $\overline \nu$ & $2.477 \pm 0.008$ & $2.479 \pm 0.004$ & $1.00 \pm 0.01$ & $2.130 \pm 0.001$ & $0.86 \pm 0.01$ \\
 & $\nu_2$ & $4.850 \pm 0.242$ & $4.848 \pm 0.014$ & $1.00 \pm 0.05$ & $3.399 \pm 0.003$ & $0.70 \pm 0.04$ \\
 & $\nu_3$ & $7.46 \pm 0.373$ & $7.084 \pm 0.046$ & $0.95 \pm 0.05$ & $3.775 \pm 0.002$ & $0.51 \pm 0.03$ \\
       
\colrule
\multirow{3}{*}{$^{235}$U($n_{\rm th}$,f)}

 & $\overline \nu$ & $2.414 \pm 0.007$ & $2.391 \pm 0.016$ & $0.99 \pm 0.01$ & $2.170 \pm 0.001$ & $0.90 \pm 0.01$ \\
 & $\nu_2$ & $4.638 \pm 0.030$ & $4.543 \pm 0.059$ & $0.98 \pm 0.01$ & $3.620 \pm 0.003$ & $0.78 \pm 0.01$ \\
 & $\nu_3$ & $6.818 \pm 0.168$ & $6.564 \pm 0.132$ & $0.96 \pm 0.03$ & $4.381 \pm 0.002$ & $0.64 \pm 0.02$ \\

\colrule
\multirow{3}{*}{$^{239}$Pu($n_{\rm th}$,f)}

 & $\overline \nu$ & $2.876 \pm 0.009$ & $2.907 \pm 0.003$ & $1.01 \pm 0.01$ & $2.649 \pm 0.001$ & $0.92 \pm 0.01$ \\
 & $\nu_2$ & $6.744 \pm 0.018$ & $6.782 \pm 0.038$ & $1.01 \pm 0.01$ & $5.544 \pm 0.005$ & $0.82 \pm 0.01$ \\
 & $\nu_3$ & $12.545 \pm 0.054$ & $12.190 \pm 0.266$ & $0.97 \pm 0.02$ & $8.736 \pm 0.002$ & $0.70 \pm 0.01$ \\

\colrule
\multirow{3}{*}{$^{241}$Pu($n_{\rm th}$,f)}

 & $\overline \nu$ & $2.93 \pm 0.11$ & $2.923 \pm 0.01$ & $1.00 \pm 0.04$ & $2.526 \pm 0.001$ & $0.86 \pm 0.03$ \\ 
 & $\nu_2$ & $6.60 \pm 0.90$ & $6.784 \pm 0.04$ & $1.03 \pm 0.14$ & $4.978 \pm 0.004$ & $0.75 \pm 0.10$ \\
 & $\nu_3$ & $12.50 \pm 3.80$ & $11.917 \pm 0.087$ & $0.95 \pm 0.29$ & $7.229 \pm 0.002$ & $0.58 \pm 0.18$ \\

\end{tabular}
\end{ruledtabular}
\end{table*}

In Table~\ref{tab:nf-photon-comparison}, $\overline N_\gamma$ and $\overline \epsilon_\gamma$ calculated with $\mathtt{FREYA}$ are compared to measured data for thermal neutron-induced fission.  Using the parameter values determined in this work, all data are reproduced within uncertainties, except for $\overline N_\gamma$ from $^{235}$U($n_{\rm th}$,f).  The values from this work have achieved improved agreement with the $\overline N_\gamma$ and $\overline \epsilon_\gamma$ data for $^{233}$U($n_{\rm th}$,f) and $^{241}$Pu($n_{\rm th}$,f) compared to the parameter values assumed in Ref.~\cite{Verbeke2018}.  In the case of $^{235}$U($n_{\rm th}$,f) and $^{239}$Pu($n_{\rm th}$,f), C/E is comparable for both sets of parameters, except for $\overline N_\gamma$ from $^{235}$U($n_{\rm th}$,f).

\begin{table*}
\caption{Same as Table~\ref{tab:sf-photon-comparison} but for thermal neutron-induced fission and the $\mathtt{FREYA}$ results using the default parameter values from Ref.~\cite{Verbeke2018}. Note that the uncertainties on the $\mathtt{FREYA}$ results using values from Ref.~\cite{Verbeke2018} are purely statistical, whereas we have, in addition to the statistical uncertainty, propagated the fit uncertainties through to our results.}
\label{tab:nf-photon-comparison}
\begin{ruledtabular}
\begin{tabular}{llccccc} 
 & & & \multicolumn{2}{c}{This work} & \multicolumn{2}{c}{Ref.~\cite{Verbeke2018}} \\ 

Reaction & Average & Literature & $\mathtt{FREYA}$ & C/E & $\mathtt{FREYA}$ & C/E \\

\colrule
\multirow{2}{*}{$^{233}$U($n_{\rm th}$,f)}

 & $\overline N_\gamma$ & $6.31 \pm 0.30$ & $6.278 \pm 0.391$ & $0.99 \pm 0.08$ & $6.493 \pm 0.002$ & $1.03 \pm 0.05$ \\
 & $\overline \epsilon_\gamma$ (MeV) & $1.06 \pm 0.07$ & $1.059 \pm 0.016$ & $1.00 \pm 0.07$ & $1.031 \pm 0.001$ & $0.97 \pm 0.06$ \\
       
\colrule
\multirow{2}{*}{$^{235}$U($n_{\rm th}$,f)}

 & $\overline N_\gamma$ & $6.51 \pm 0.30$ & $ 7.853 \pm 0.068$ & $1.21 \pm 0.06$ & $6.372 \pm 0.002$ & $0.98 \pm 0.05$ \\
 & $\overline \epsilon_\gamma$ (MeV) & $0.99 \pm 0.07$ & $0.949 \pm 0.002$ & $0.96 \pm 0.07$ & $0.989 \pm 0.001$ & $1.0 \pm 0.07$ \\

\colrule
\multirow{2}{*}{$^{239}$Pu($n_{\rm th}$,f)}

 & $\overline N_\gamma$ & $6.88 \pm 0.35$ & $6.648 \pm 0.739$ & $0.97 \pm 0.12$ & $6.766 \pm 0.002$ & $0.98 \pm 0.05$ \\
 & $\overline \epsilon_\gamma$ (MeV) & $0.98 \pm 0.07$ & $1.022 \pm 0.025$ & $1.04 \pm 0.08$ & $1.007 \pm 0.001$ & $1.03 \pm 0.07$ \\

\colrule
\multirow{2}{*}{$^{241}$Pu($n_{\rm th}$,f)}

 & $\overline N_\gamma$ & $7.51 \pm 0.04$ & $7.511 \pm 0.003$ & $1.00 \pm 0.01$ & $6.774 \pm 0.002$ & $0.90 \pm 0.01$ \\
 & $\overline \epsilon_\gamma$ (MeV) & $0.86 \pm 0.22$ & $0.910 \pm 0.016$ & $1.06 \pm 0.27$ & $0.958 \pm 0.001$ & $1.11 \pm 0.29$ \\

\end{tabular}
\end{ruledtabular}
\end{table*}

\begin{figure*}
    \includegraphics[width=1.0\linewidth]{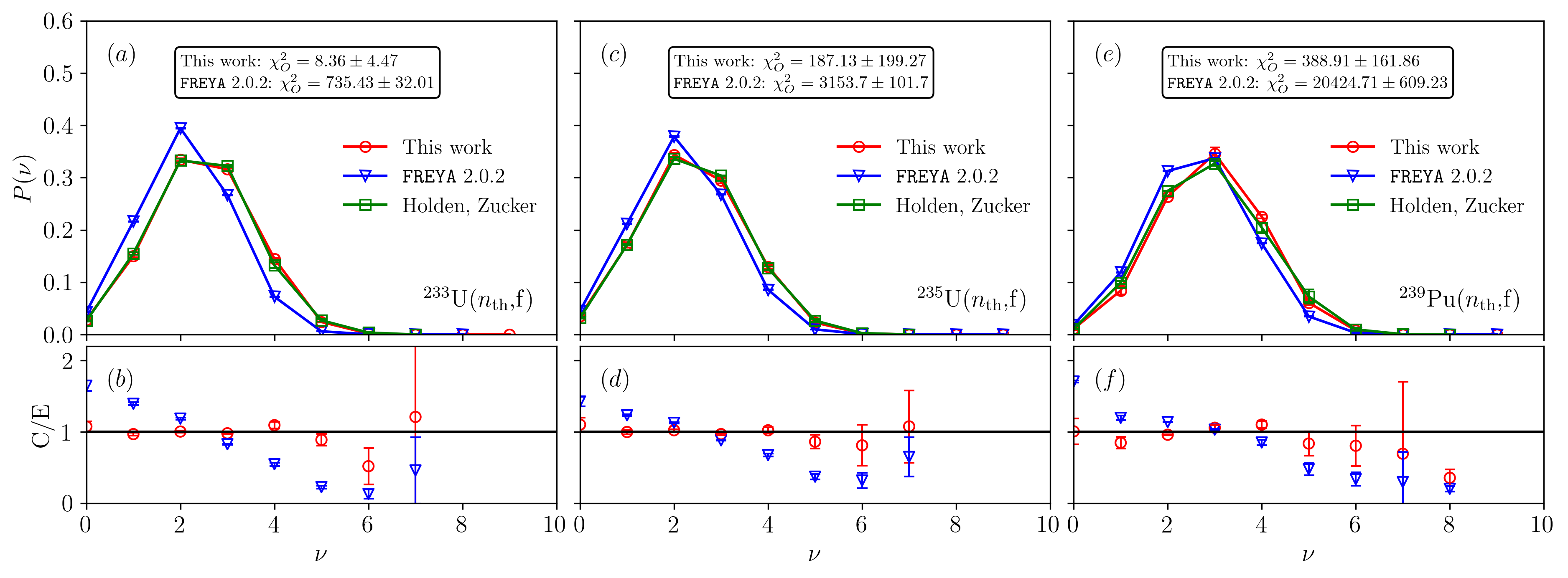}
    \caption{(a), (c), and (e) The neutron multiplicity distributions for $^{233}$U($n_{\rm th}$,f), $^{235}$U($n_{\rm th}$,f), and $^{239}$Pu($n_{\rm th}$,f) using parameter values from this work and Ref.~\cite{Verbeke2018} as well as the Holden-Zucker evaluation~\cite{Holden-Zucker1988}. (b), (d), and (f) Ratio of $\mathtt{FREYA}$ calculations to the evaluated results.}
    \label{fig:nf_P(nu)}
\end{figure*}

As can be seen in Fig.~\ref{fig:nf_P(nu)}, the neutron multiplicity distributions, $P(\nu)$, from the Holden-Zucker evaluations are reproduced almost exactly for $^{233}$U($n_{\rm th}$,f), $^{235}$U($n_{\rm th}$,f), and $^{239}$Pu($n_{\rm th}$,f) using the parameter values determined in this work. At high neutron multiplicities, where the $\mathtt{FREYA}$ results deviate significantly from the evaluation, the uncertainties on C/E tend to be compatible with unity. The $\chi_O^2$ has been significantly reduced relative to the $\mathtt{FREYA}$ results based on the parameter values of Ref.~\cite{Verbeke2018}, between one and two orders of magnitude. 


\begin{figure*}
    \includegraphics[width=1.0\linewidth]{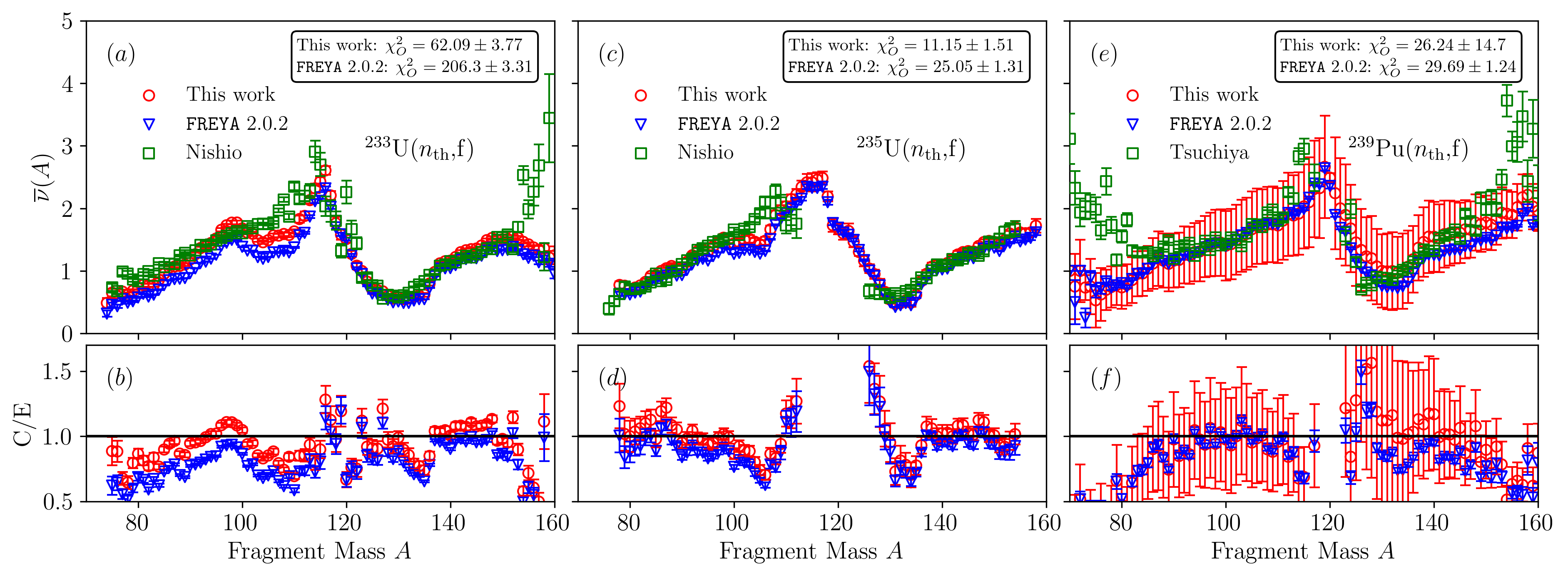}
    \caption{(a), (c), and (e) The average neutron multiplicity as a function of fragment mass using parameter values from this work and Ref.~\cite{Verbeke2018} along with experimental data on $^{233}$U($n_{\rm th}$,f)~\cite{Nishio1998:233U}, $^{235}$U($n_{\rm th}$,f)~\cite{Nishio1998:235U}, and $^{239}$Pu($n_{\rm th}$,f)~\cite{Tsuchiya2000} respectively. Note that the large uncertainties on our new results in (e) and (f) reflect the large uncertainty on the $x$ parameter for $^{239}$Pu($n_{\rm th}$,f), see Table~\ref{tab:nfparameters}, which directly affects $\nu(A)$. (b), (d), and (f) Ratio of $\mathtt{FREYA}$ calculations to the experimental results.}
    \label{fig:nf_nu(A)}
\end{figure*}

In Fig.~\ref{fig:nf_nu(A)}, the average neutron multiplicity as a function of fragment mass, $\overline \nu(A)$, calculated with $\mathtt{FREYA}$ is compared to experimental data from $^{233}$U($n_{\rm th}$,f)~\cite{Nishio1998:233U}, $^{235}$U($n_{\rm th}$,f)~\cite{Nishio1998:235U}, and $^{239}$Pu($n_{\rm th}$,f)~\cite{Tsuchiya2000}.  The $^{233}$U($n_{\rm th}$,f) data are reproduced reasonably well outside the region near symmetric fission. Compared to $^{235}$U($n_{\rm th}$,f), there are dips in the $\mathtt{FREYA}$ calculation near $A = 105$ and 135 that are not present in the Nishio measurement. Aside from these features, the experimental data is reproduced well. Finally, the $\mathtt{FREYA}$ calculation for $^{239}$Pu($n_{\rm th}$,f) diverges from measurement at the lowest and highest fragment masses, but ${\rm C/E} \sim 1$ within the uncertainties in the mass range $80 < A < 150$. Furthermore, the Tsuchiya data outside this mass range appear to be outliers due to the large uncertainties and noise present in the data. Note that the large uncertainties on the results using the parameters from this work reflect the large uncertainty on the $x$ parameter for $^{239}$Pu($n_{\rm th}$,f), see Table~\ref{tab:nfparameters}, which directly affects $\overline \nu(A)$. Thus, aside from $^{239}$Pu($n_{\rm th}$,f), the $\chi_O^2$ for $\overline \nu(A)$ calculated with our optimized parameters is reduced relative to that calculated using the values assumed in Ref.~\cite{Verbeke2018}. 


\begin{figure*}
    \includegraphics[width=1.0\linewidth]{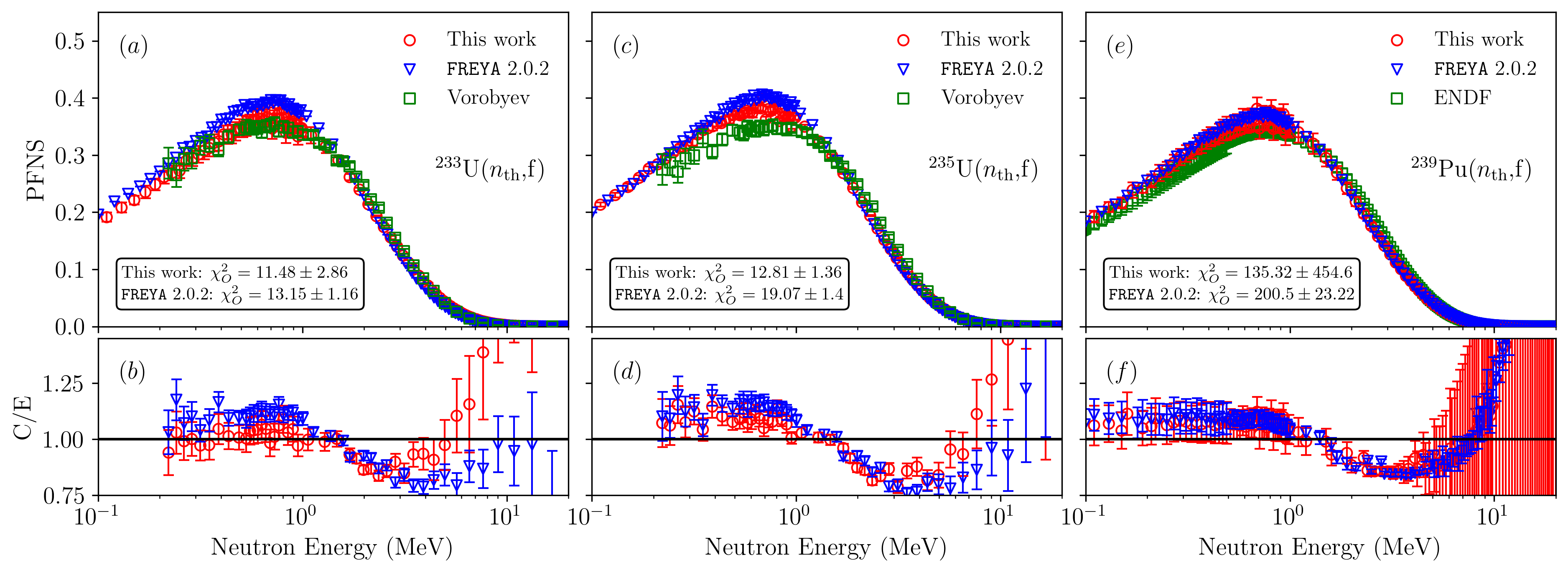}
    \caption{(a), (c), and (e) The PFNS for $^{233}$U($n_{\rm th}$,f), $^{235}$U($n_{\rm th}$,f), and $^{239}$Pu($n_{\rm th}$,f) using parameter values from this work and Ref.~\cite{Verbeke2018} as well as measurements from Ref.~\cite{Vorobyev2016} for $^{233,235}$U and the $^{239}$Pu ENDF/B-VIII.0 evaluation~\cite{Brown2018}. (b), (d), and (f) Ratio of calculated ($\mathtt{FREYA}$) to experimental or evaluated results. Note the large uncertainties on the C/E ratio for $^{239}$Pu from this work in (f) is a result of propagating the optimized parameter uncertainties.}
    \label{fig:nf_pfns}
\end{figure*}

$\mathtt{FREYA}$ reproduces the PFNS data employed in our fits well for all thermal neutron-induced fission reactions, as seen in Fig.~\ref{fig:nf_pfns}. Employing the parameters determined in this work, the $\chi_O^2$ for $^{235}$U($n_{\rm th}$,f) is reduced relative to that obtained with the values of Ref.~\cite{Verbeke2018}, while for $^{233}$U($n_{\rm th}$,f) and $^{239}$Pu($n_{\rm th}$,f) the $\chi_O^2$ are consistent within uncertainties. Note that the relative uncertainties on the ENDF/B-VIII.0 evaluation of $^{239}$Pu($n_{\rm th}$,f) abruptly decrease near 0.5~MeV outgoing neutron energy, as mentioned in Sec.~\ref{sec:data}.  In all cases, C/E tends to be slightly greater than unity for outgoing neutron energies below $\sim1$~MeV, before dropping below unity above this energy until $\sim10$~MeV with ${\rm C/E}>1$ for the highest neutron energies.  At these higher neutron energies, where $\mathtt{FREYA}$ begins to diverge from the data, the uncertainties on C/E are still compatible with unity for most data points.


\subsection{Model predictions}

The optimized parameters presented in Tables~\ref{tab:sfparameters} and~\ref{tab:nfparameters} are now used to predict fission observables that were not included in our fits. We present direct comparisons as well as C/E ratios. As described in the previous subsection, the uncertainties on the $\mathtt{FREYA}$ results reflect the  statistical uncertainties on the $\mathtt{FREYA}$ calculation as well as the propagation of the parameter uncertainties from the fits.  The $\mathtt{FREYA}$ results using parameters from Ref.~\cite{VanDyke2019} for spontaneous fission or Ref.~\cite{Verbeke2018} for thermal neutron-induced fission are also shown. These $\mathtt{FREYA}$ results include statistical uncertainties only.

\begin{figure}
    \includegraphics[width=1.0\linewidth]{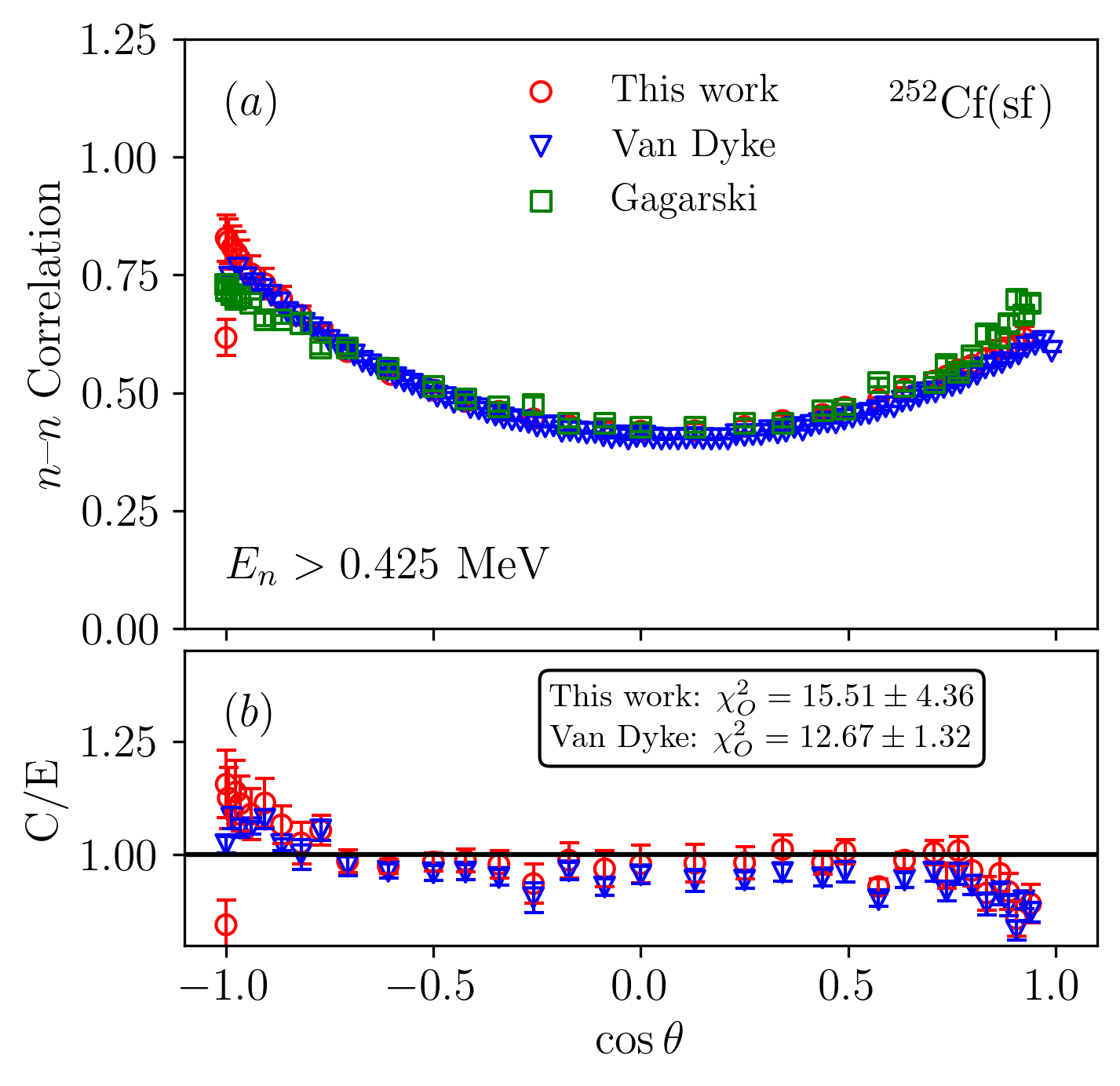}
    \caption{(a) Neutron-neutron correlations for $^{252}$Cf(sf) using parameter values from this work and Ref.~\cite{VanDyke2019} as well as a measurement from Ref.~\cite{Gagarski2008} not included in our fit. (b) Ratio of $\mathtt{FREYA}$ calculations to the measurement.}
    \label{fig:Cf252sf_n-n}
\end{figure}

Reference~\cite{Gagarski2008} measured $n$--$n$ correlations in $^{252}$Cf(sf). As discussed in Sec.~\ref{sec:data}, we have chosen not to include these data in our $^{252}$Cf(sf) fit. We nonetheless compare the $\mathtt{FREYA}$ result to the data in Fig.~\ref{fig:Cf252sf_n-n} for neutron energies, $E_n$, greater than 0.425~MeV. Even though these data were not included in our fit, the $\mathtt{FREYA}$ results are in good agreement with them.  Additionally, the $\chi_O^2$ obtained using the parameters determined in this work and those from Ref.~\cite{VanDyke2019} agree within uncertainties. We note that for higher neutron energy thresholds (not shown), the $\mathtt{FREYA}$ results produce equivalent agreement with the Gagarski data ($|{\rm C/E}-1|<0.20$ for all neutron angles). The excellent agreement in Fig.~\ref{fig:Cf252sf_n-n}, when these data were not included, either here or in Ref.~\cite{VanDyke2019}, supports our finding that substituting $\overline \nu(A)$ data from Ref.~\cite{Dushin2004} for $n$--$n$ correlation data has minimal effect on the fit results for $^{252}$Cf(sf), as discussed in Sec.~\ref{sec:data}.  This result also supports the general observation that, if $\nu(A)$ data are not available to use to optimize the $x$ parameter, $n$--$n$ correlation data, if available, can be used as a proxy.

\begin{figure}
    \includegraphics[width=1.0\linewidth]{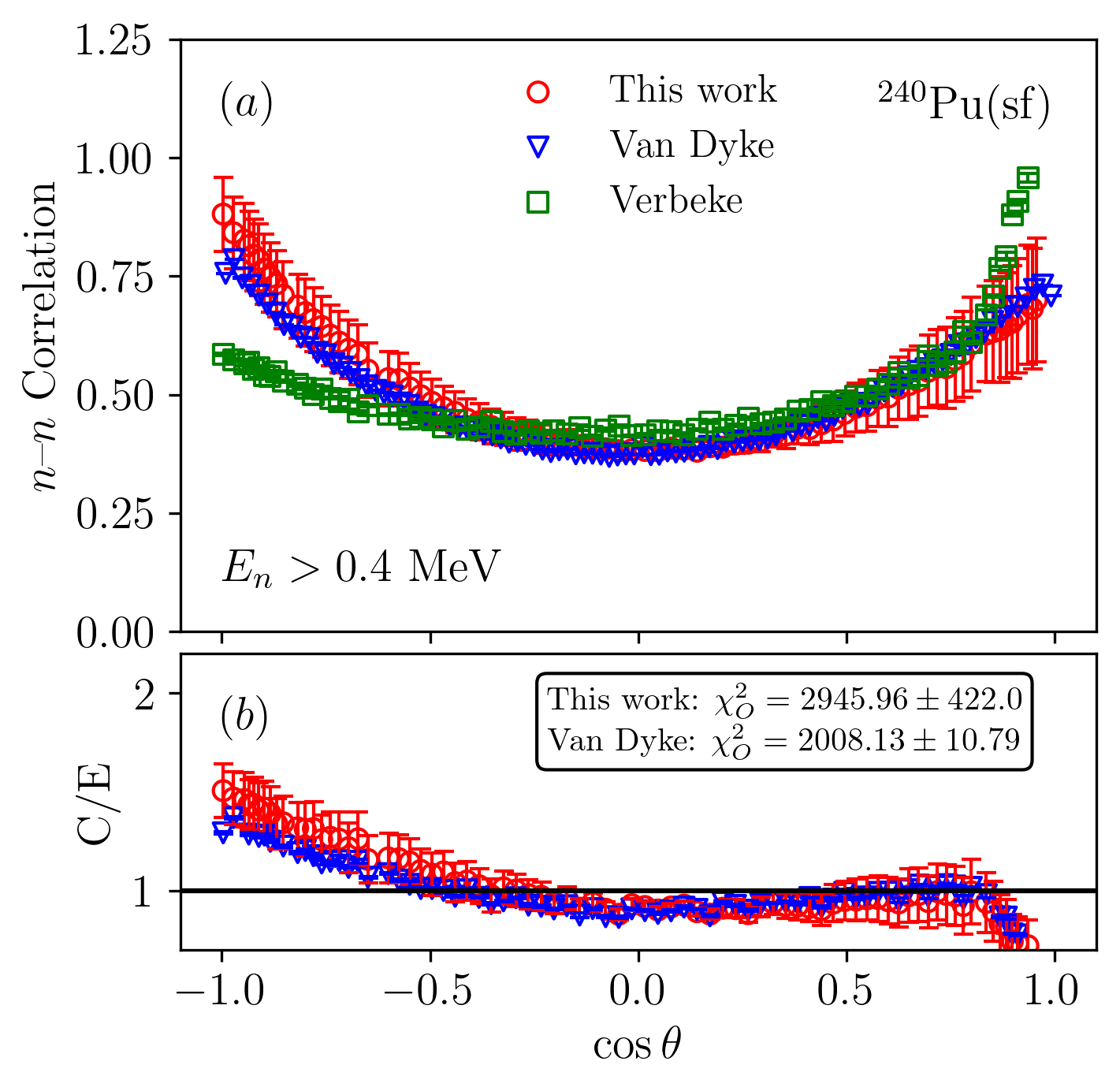}
    \caption{Same as Fig.~\ref{fig:Cf252sf_n-n} but for $^{240}$Pu(sf) and a measurement from Ref.~\cite{Verbeke2018nn} not included in our fit.}
    \label{fig:Pu240sf_n-n}
\end{figure}

Although we chose not include the $n$--$n$ correlation data from Ref.~\cite{Verbeke2018nn} in our fit for $^{240}$Pu(sf),  we compare our $\mathtt{FREYA}$ result to these data in Fig.~\ref{fig:Pu240sf_n-n} for $E_n >0.40$~MeV (the lowest energy threshold for this measurement). The $\mathtt{FREYA}$ results with the parameters obtained here and with those obtained in Ref.~\cite{VanDyke2019} predict a higher $n$--$n$ correlation when one neutron from each fragment is used in the correlation. Although not shown here, as the neutron energy threshold is increased, the agreement between the $\mathtt{FREYA}$ predictions and the data improve, especially at $\theta = 180^\circ$. We show the results with the poorest agreement with $\mathtt{FREYA}$ to more closely match the energy threshold of the Gagarski data shown in Fig.~\ref{fig:Cf252sf_n-n}.  

We note that the uncertainties are larger with the parameters obtained here than in Ref.~\cite{VanDyke2019} because our uncertainty on $x$ is larger for $^{240}$Pu(sf) in this case.  In addition, the overall $\chi_O^2$ determined in this work is greater than that obtained with the parameters of Ref.~\cite{VanDyke2019}.  As discussed in Sec.~\ref{sec:data}, including the $n$--$n$ data in the fit for $^{240}$Pu(sf) degraded the quality of agreement with the PFNS measurement from Ref.~\cite{Gerasimenko2002}.  Since these two data sets are in tension with one another, it is therefore not surprising that the parameter values from Ref.~\cite{VanDyke2019} better reproduce the $n$--$n$ data, considering that our parameter values better reproduce the PFNS, see Fig.~\ref{fig:Pu240sf_PFNS}. 


\begin{figure*}
    \includegraphics[width=1.0\linewidth]{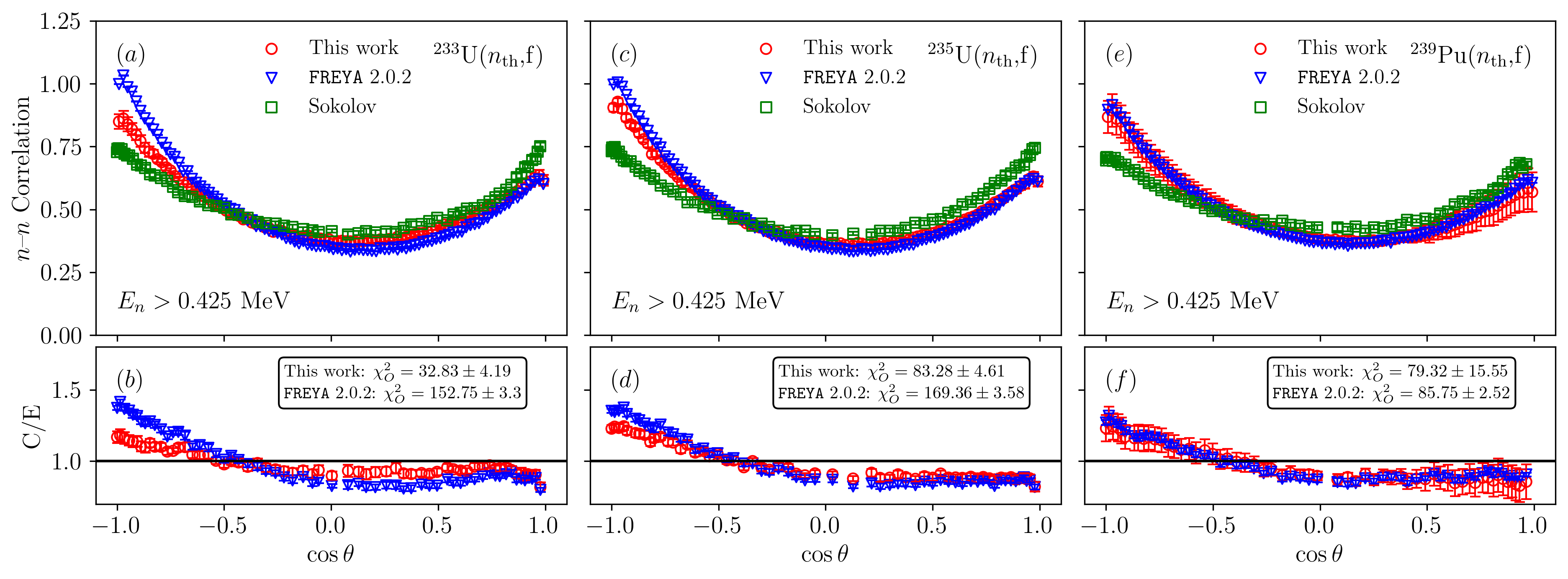}
    \caption{(a), (c), and (e) The $n$--$n$ correlations for $^{233}$U($n_{\rm th}$,f), $^{235}$U($n_{\rm th}$,f), and $^{239}$Pu($n_{\rm th}$,f) using parameter values from this work and Ref.~\cite{Verbeke2018} as well as measurements from Ref.~\cite{Sokolov2010}. (b), (d), and (f) Ratio of $\mathtt{FREYA}$ calculations to the experimental results.}
    \label{fig:nf_nn}
\end{figure*}

In Fig.~\ref{fig:nf_nn}, the neutron-neutron correlations calculated from the $\mathtt{FREYA}$ output for the respective reactions $^{233}$U($n_{\rm th}$,f), $^{235}$U($n_{\rm th}$,f), and $^{239}$Pu($n_{\rm th}$,f) are compared to measured data from Ref.~\cite{Sokolov2010}.  These data were not included in the fits for thermal neutron-induced fission. The $\chi_O^2$ achieved with the parameters determined in this work for $^{233}$U($n_{\rm th}$,f) and $^{235}$U($n_{\rm th}$,f) have been significantly improved relative to that obtained with the values of Ref.~\cite{Verbeke2018}, whereas the $\chi_O^2$ agree within uncertainties for $^{239}$Pu($n_{\rm th}$,f).  This work has therefore generally improved the predictive capabilities of $\mathtt{FREYA}$.  Moreover, these comparisons indicate that the excitation energy sharing encapsulated in the $x$ parameter determines the shape of the $n$--$n$ correlations, as discussed in Ref.~\cite{Vogt:2014}.


\begin{figure*}
    \includegraphics[width=1.0\linewidth]{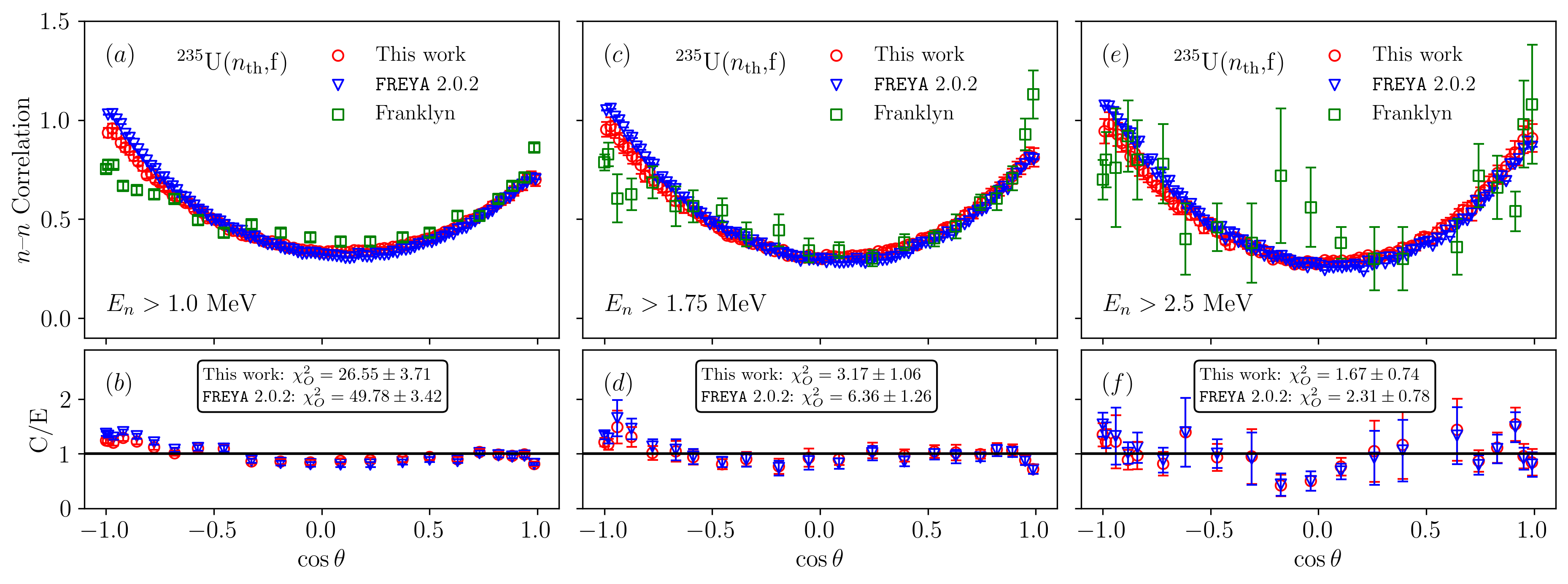}
    \caption{(a), (c), and (e) The $n$--$n$ correlations for $^{235}$U($n_{\rm th}$,f) for varying neutron energy thresholds using parameter values from this work and Ref.~\cite{Verbeke2018} as well as measurements from Ref.~\cite{Franklyn1978}. (b), (d), and (f) Ratio of $\mathtt{FREYA}$ calculations to the experimental results.}
    \label{fig:U235nf_nn}
\end{figure*}

Figure~\ref{fig:U235nf_nn} compares the $n$--$n$ correlations calculated with $\mathtt{FREYA}$ for $^{235}$U($n_{\rm th}$,f) measurements at three different neutron kinetic energy thresholds~\cite{Franklyn1978}. Except for the highest threshold, $E_n>2.5$~MeV, where the measured uncertainties are large, the $\chi_O^2$ calculated with our optimized parameters is significantly smaller than that obtained using the assumed values from Ref.~\cite{Verbeke2018}.  At the highest threshold, the $\chi_O^2$ agree within uncertainties.  The use of neutron energy thresholds in $n$--$n$ correlations samples neutrons from different regions of the PFNS.  Thus a more accurate reproduction of the PFNS should lead to more accurate predictions of $n$--$n$ correlations over a range of energy thresholds.  Indeed, the parameter values found here better reproduce the PFNS of $^{235}$U($n_{\rm th}$,f), see Fig.~\ref{fig:nf_pfns}.

\section{Conclusions}
\label{sec:conclusion}
We have improved the parameter determinations for spontaneous fission relative to Ref.~\cite{VanDyke2019} by using a genetic algorithm. We found that, in many cases, the genetic algorithm provided superior results over those with simulated annealing. We then extended our genetic algorithm to determine the $\mathtt{FREYA}$ parameters for thermal neutron-induced fission, capturing the isotope dependence of these parameters for the first time. The isotope-specific parameter values determined in this work have significantly improved the agreement with available evaluated and measured data. Consequently, the predictive capabilities of $\mathtt{FREYA}$ for thermal neutron-induced fission have also been improved. In future work, we will apply the fitting procedure we have developed here to neutron-induced fission for incident neutron energies up to 20~MeV to determine the energy dependence of the $\mathtt{FREYA}$ parameters. 

For example, as nuclear shell effects wash out at higher incident neutron energies, the ``sawtooth'' in $\overline \nu(A)$ should become less pronounced, resulting in an energy dependence of the $x$ parameter. In addition,  increasing the excitation energy at scission should increase the rotational energy of the fragments, resulting in higher fragment spins which will produce an energy dependent $c_S$. Our future studies of the energy dependence of the $\mathtt{FREYA}$ parameters will provide insights into the energy dependence of particle emission during fission.

\section{Acknowledgements}
We wish to acknowledge helpful conversations with Jørgen Randrup, Jerome Verbeke, Jackson Van Dyke, and Nathan Giha. The computational work was done on the Great Lakes cluster with a computing allowance provided by the University of Michigan Advanced Research Computing. The work of R. Vogt is supported by Lawrence Livermore National Laboratory under Contract DE-AC52-07NA27344. This work was supported by the Office of Defense Nuclear Nonproliferation Research and Development within the U.S. Department of Energy’s National Nuclear Security Administration.

\appendix
\renewcommand{\thefigure}{\thesection\arabic{figure}}
\renewcommand{\thetable}{\thesection\arabic{table}}

\counterwithin{figure}{section}
\counterwithin{table}{section}

\section{Comparison to photon observables in thermal neutron-induced fission}
\label{sec:A}

Here we compile comparisons of the $\mathtt{FREYA}$ results with the photon data available for $^{233,235}$U($n_{\rm th}$,f) and $^{239,241}$Pu($n_{\rm th}$,f). 

\begin{figure*}
    \includegraphics[width=1.0\linewidth]{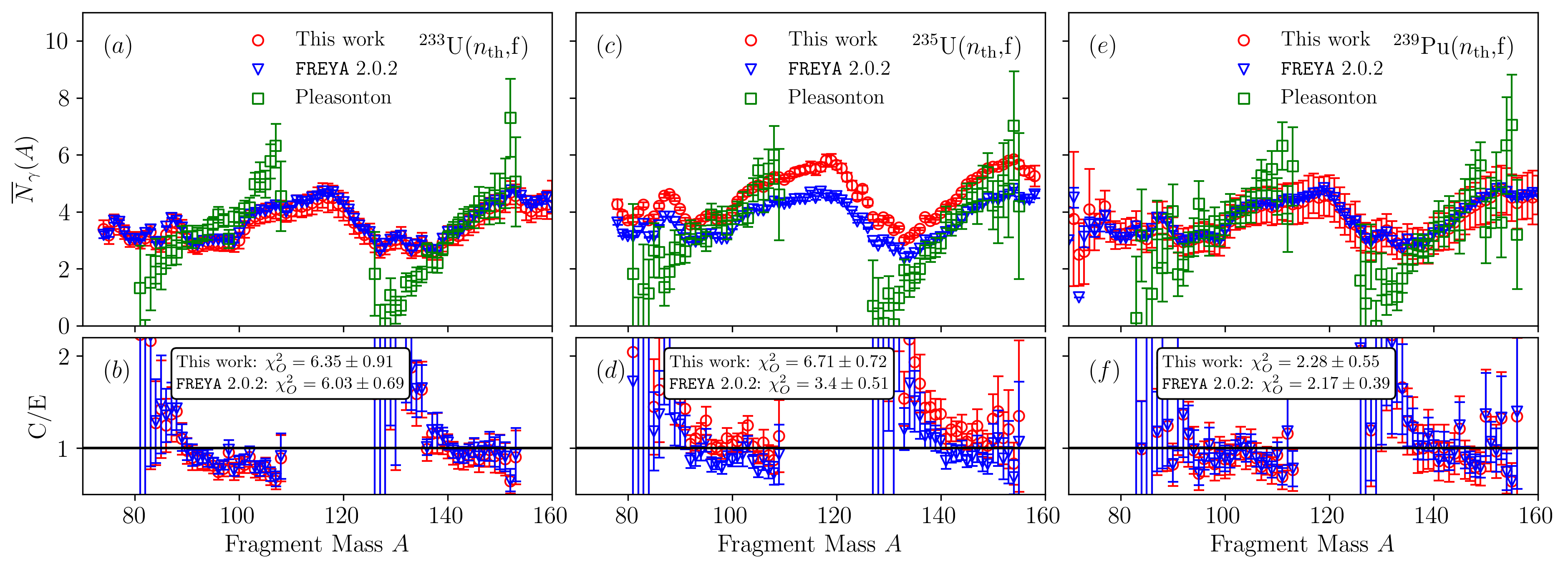}
    \caption{(a), (c), and (e) The average photon multiplicity as a function of fragment mass for $^{233}$U($n_{\rm th}$,f), $^{235}$U($n_{\rm th}$,f), and $^{239}$Pu($n_{\rm th}$,f) using parameter values from this work and Ref.~\cite{Verbeke2018} as well as measurements from Refs.~\cite{Pleasonton1973,Pleasonton1972}. (b), (d), and (f) Ratio of $\mathtt{FREYA}$ calculations to the experimental results.}
    \label{fig:nf_m(A)}
\end{figure*}

\begin{figure*}
    \includegraphics[width=1.0\linewidth]{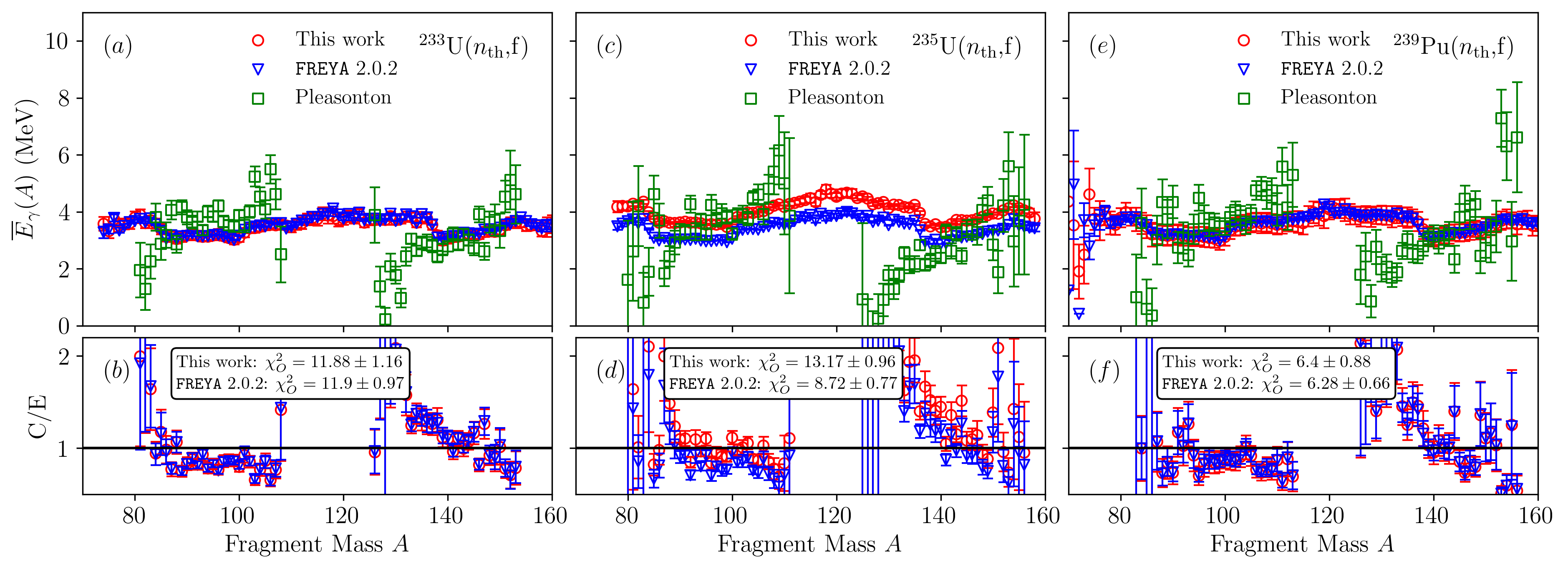}
    \caption{(a), (c), and (e) The average total photon energy as a function of fragment mass for $^{233}$U($n_{\rm th}$,f), $^{235}$U($n_{\rm th}$,f), and $^{239}$Pu($n_{\rm th}$,f) using parameter values from this work and Ref.~\cite{Verbeke2018} as well as measurements from Refs.~\cite{Pleasonton1973,Pleasonton1972}. (b), (d), and (f) Ratio of $\mathtt{FREYA}$ calculations to the experimental results.}
    \label{fig:nf_totEg(A)}
\end{figure*}

\begin{figure}[htbp]
    \includegraphics[width=1.0\linewidth]{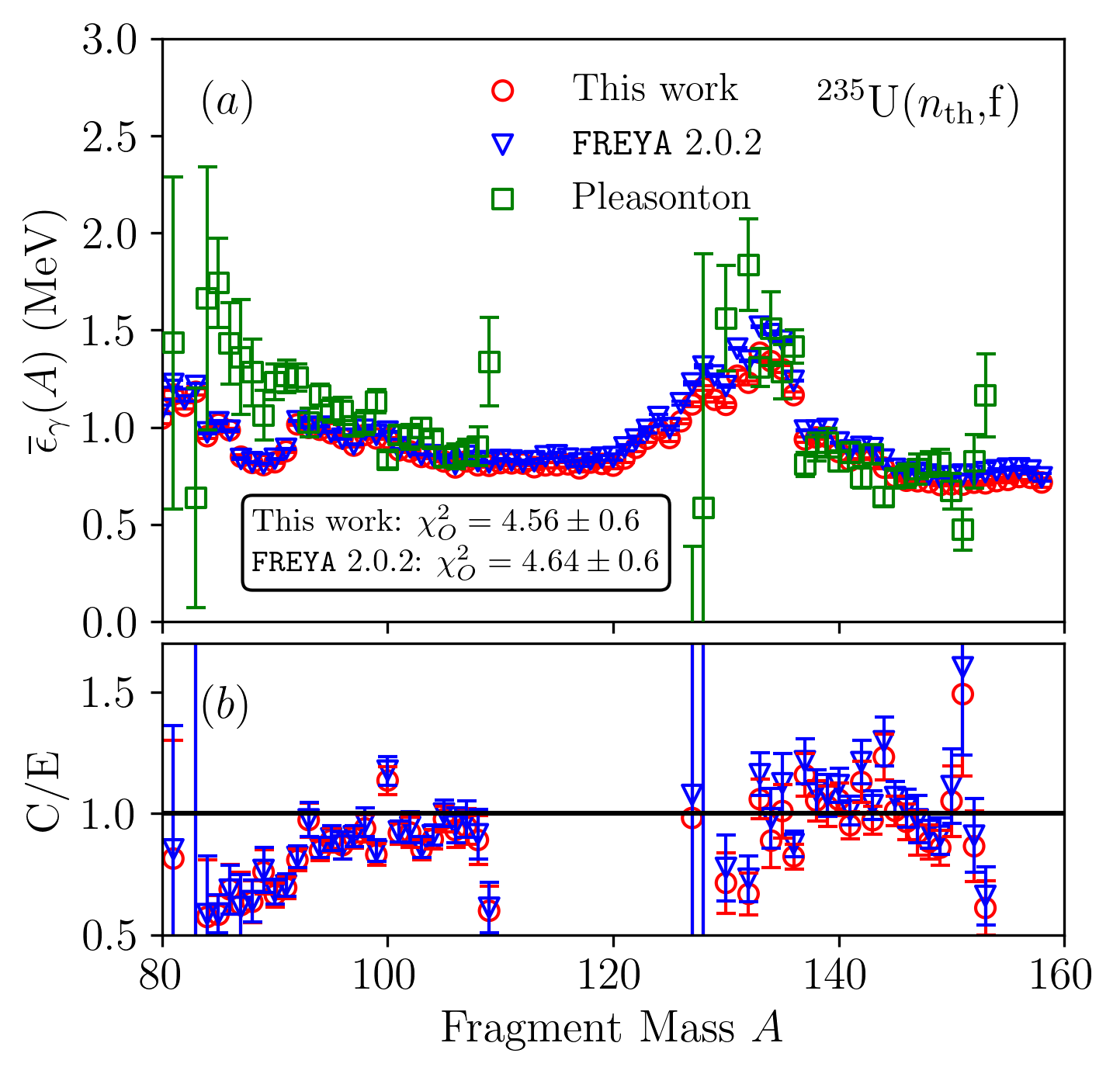}
    \caption{(a) The average photon energy as a function of fragment mass for $^{235}$U($n_{\rm th}$,f) using parameter values from this work and Ref.~\cite{Verbeke2018} compared to experimental data from Ref.~\cite{Pleasonton1972}. (b) Ratio of $\mathtt{FREYA}$ calculations to the measurement.}
    \label{fig:U235nf_Eg_A}
\end{figure}

\begin{figure}[htbp]
    \includegraphics[width=1.0\linewidth]{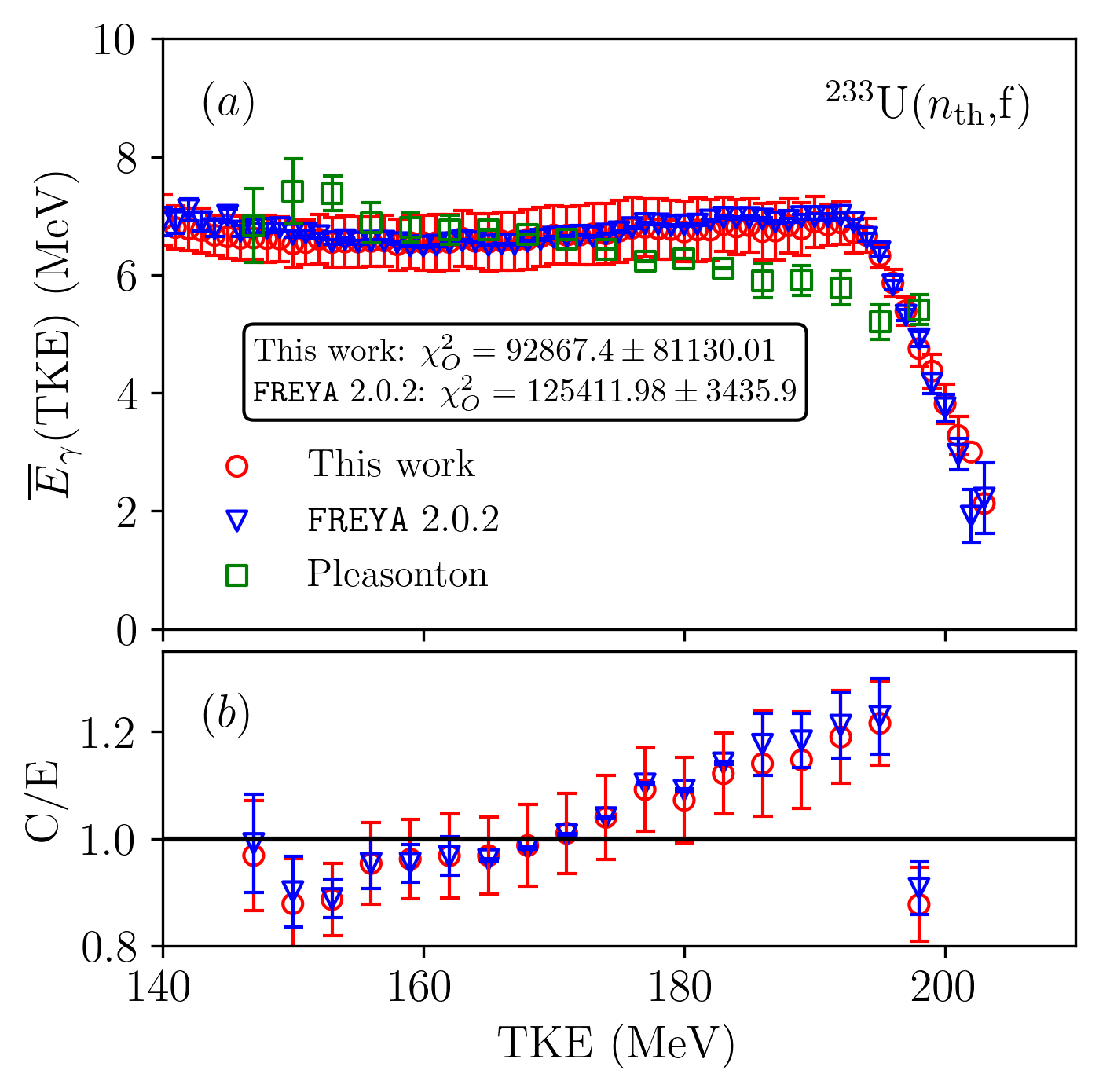}
    \caption{(a) The average total photon energy as a function of TKE for $^{233}$U($n_{\rm th}$,f) using parameter values from this work and Ref.~\cite{Verbeke2018} compared to experimental data from Ref.~\cite{Pleasonton1973}. As discussed in Sec.~\ref{sec:data}, this data set was not included in the fit. (b) Ratio of $\mathtt{FREYA}$ calculations to the experimental results.}
\label{fig:U233nf_totEg_TKE}
\end{figure}

\begin{figure}[htbp]
    \includegraphics[width=1.0\linewidth]{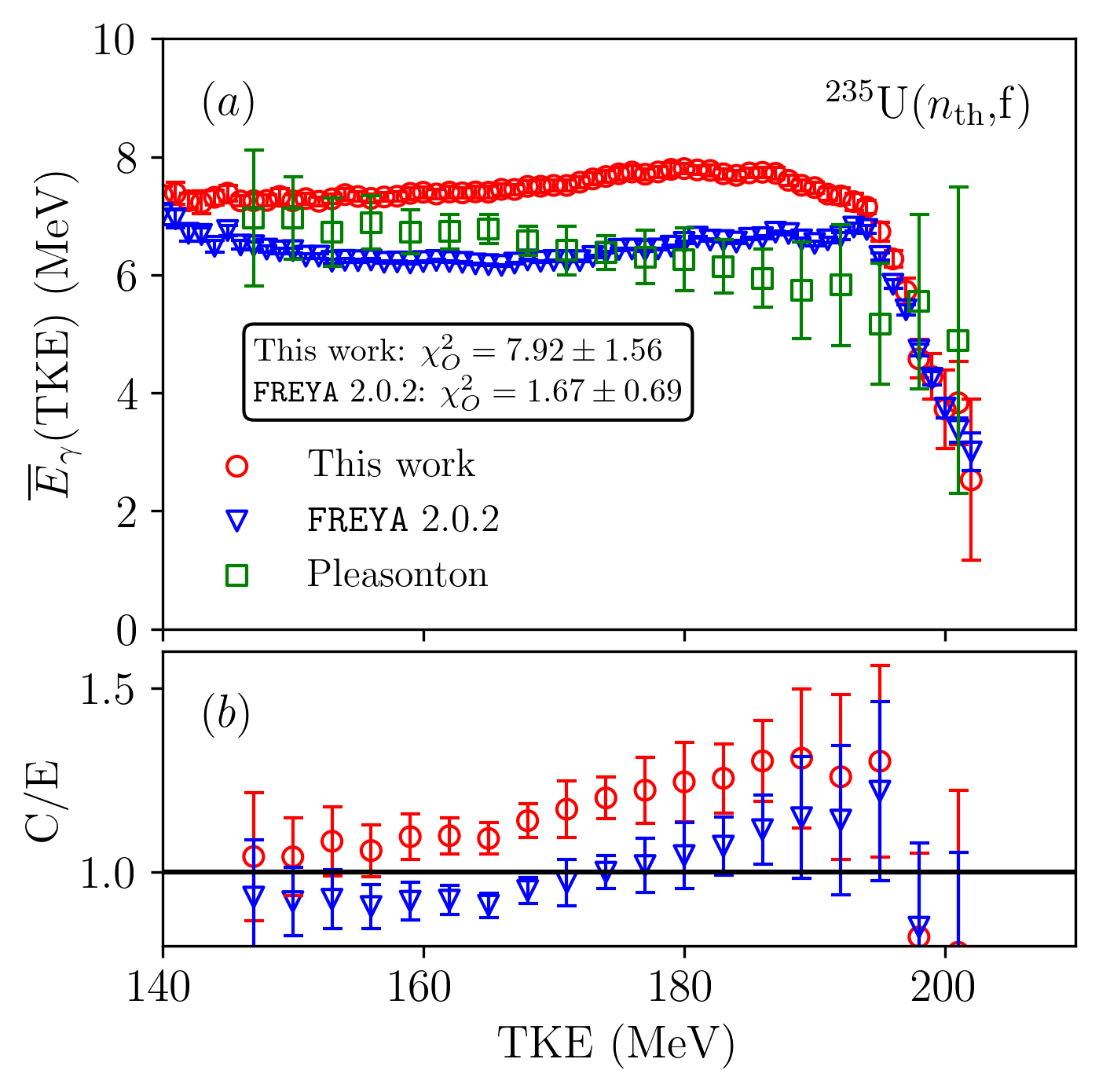}
    \caption{Same as Fig.~\ref{fig:U233nf_totEg_TKE} but for $^{235}$U($n_{\rm th}$,f) and experimental data from Ref.~\cite{Pleasonton1972}.}
\label{fig:U235nf_totEg_TKE}
\end{figure}

Pleasonton~\cite{Pleasonton1972,Pleasonton1973} took photon data on $^{233,235}$U and $^{239}$Pu correlated with the mass and total kinetic energy of the fragments: $\overline N_\gamma (A)$,  $\overline \epsilon_\gamma(A)$, $\overline E_\gamma (A)$, and $\overline E_\gamma({\rm TKE})$. As discussed in Sec.~\ref{sec:data}, these data generally have large relative uncertainties and, as a result, have a rather minimal effect on the fits. While we retained these data in the fits, their lack of influence is apparent by the overall poor agreement between the $\mathtt{FREYA}$ results and the data in Figs.~\ref{fig:nf_m(A)}--\ref{fig:U235nf_totEg_TKE}.  The differences in the two calculations are largest for $^{235}$U because our fit gives a significantly greater value of $c_S$ for this isotope, $c_S \sim 1.5$, while the original value is $\sim 0.9$.

As observed in Ref.~\cite{Vogt:2017}, the mass dependent photon observables $\overline N_\gamma(A)$ and $\overline E_\gamma (A)$ in Figs.~\ref{fig:nf_m(A)} and \ref{fig:nf_totEg(A)} appear to emulate the sawtooth shape observed for neutron emission, $\overline \nu(A)$, particularly for $\overline N_\gamma(A)$ where the photon multiplicity increases with mass for both the light and heavy fragment regions. This increase is less pronounced for the total photon energy, $\overline E_\gamma(A)$. On the other hand, the $\mathtt{FREYA}$ results exhibit a weak $A$ dependence and are, in fact, almost independent of $A$.  Despite not matching the $A$ dependence of either the photon multiplicity or the total photon energy, the energy per photon, shown in Fig.~\ref{fig:U235nf_Eg_A}, is generally well reproduced.  Note the enhancement of $\overline \epsilon_\gamma(A)$ near the doubly-magic shell closure at $A_H \sim 132$.

We note that this sawtooth-like shape was also inferred for the fragment spins as a function of mass by Wilson {\it et al.}~\cite{Wilson:2021nlm}.  Standard $\mathtt{FREYA}$ did not reproduce this behavior and was similarly flat as a function of fragment mass, see Ref.~\cite{Randrup:2021hax}.  This behavior is because standard $\mathtt{FREYA}$ assumes that the fragment moments of inertia are simply the rigid body moments of inertia, scaled by a constant factor.   When this rigid body value was replaced by the ground state deformations, away from the near-spherical nuclei at $A_H \sim 132$, $\mathtt{FREYA}$ could better reproduce the measured fragment spin as a function of mass~\cite{Randrup:2021hax}.  (It is worth noting, however, that the fragment shapes may be expected to be distorted from their ground state shapes at scission.  However, by the time photon emission begins, the fragments may well have returned to near their ground state deformations.)  The effect of this change in the fragment moment of inertia has not been compared to the photon quantities shown here but could be expected to produce a similar change in shape.  A similar change of the $A$ dependence may be produced by improving the description of the level densities, using the microscopic level densities instead of the Ignatyuk parameterization of the Fermi gas level density as is currently employed in $\mathtt{FREYA}$.  See Ref.~\cite{Albertsson} for details.  Further exploration of the effects of fragment deformation or microscopic level densities on these results could be the subject of future work.

Pleasonton also measured $\overline E_\gamma({\rm TKE})$ for $^{233,235}$U.  The data show a slight decrease with increasing TKE, albeit with rather large uncertainties for $^{235}$U, as seen in Figs.~\ref{fig:U233nf_totEg_TKE} and \ref{fig:U235nf_totEg_TKE}.  On the other hand, the $\mathtt{FREYA}$ results are effectively independent of TKE until ${\rm TKE} > 195$~MeV.  It is unclear whether or not a modified moment of inertia, as discussed above, might affect the TKE dependence of the calculation.

\begin{figure}
    \includegraphics[width=1.0\linewidth]{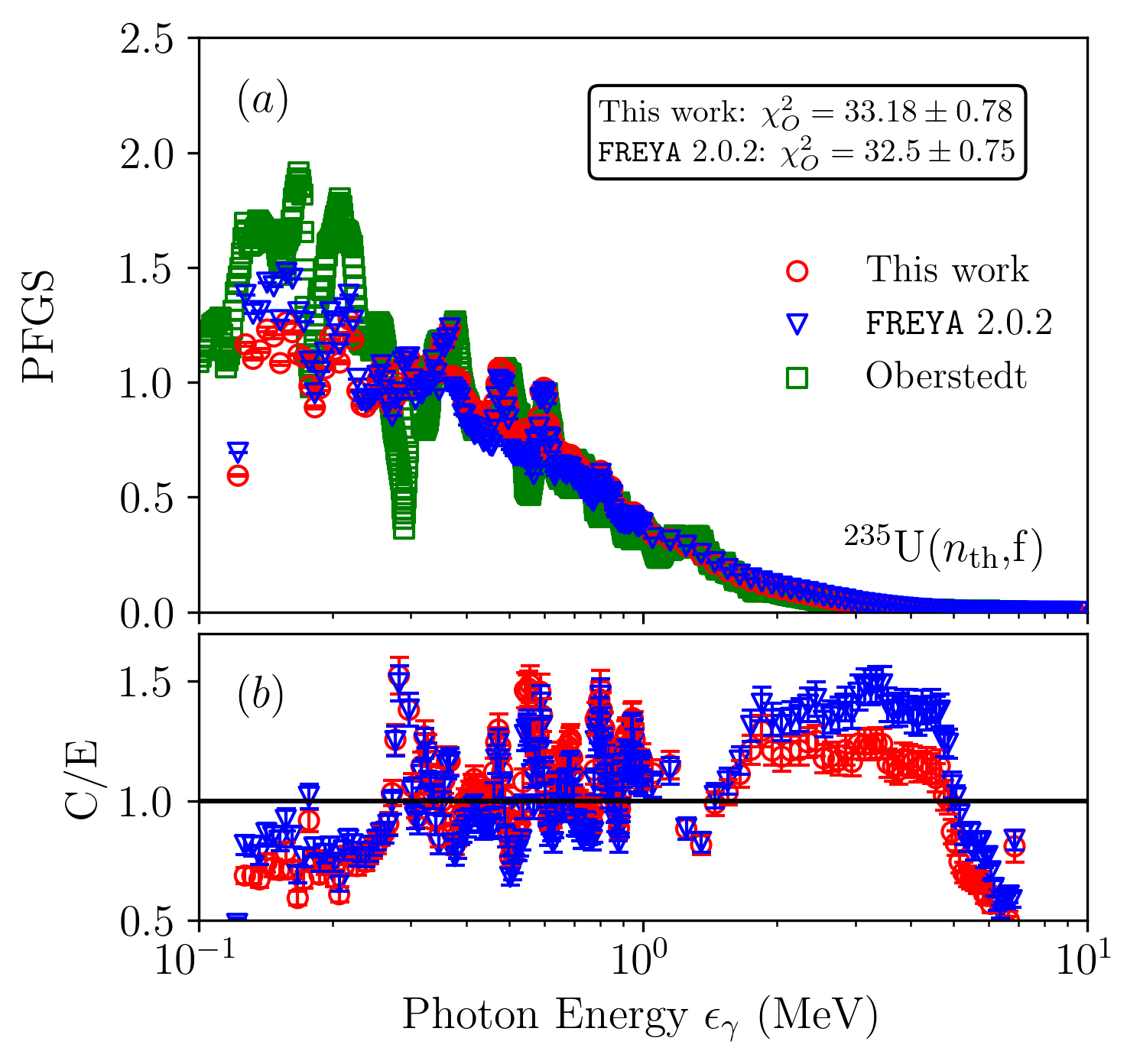}
\caption{(a) Photon energy spectrum for $^{235}$U($n_{\rm th}$,f) using parameter values from this work and Ref.~\cite{Verbeke2018} as well as experimental data from Ref.~\cite{Oberstedt2013}. Note the logarithmic scale on the $x$-axis. (b) Ratio of $\mathtt{FREYA}$ calculations to the experimental results.}
\label{fig:U235nf_PFGS}
\end{figure}

\begin{figure}
    \includegraphics[width=1.0\linewidth]{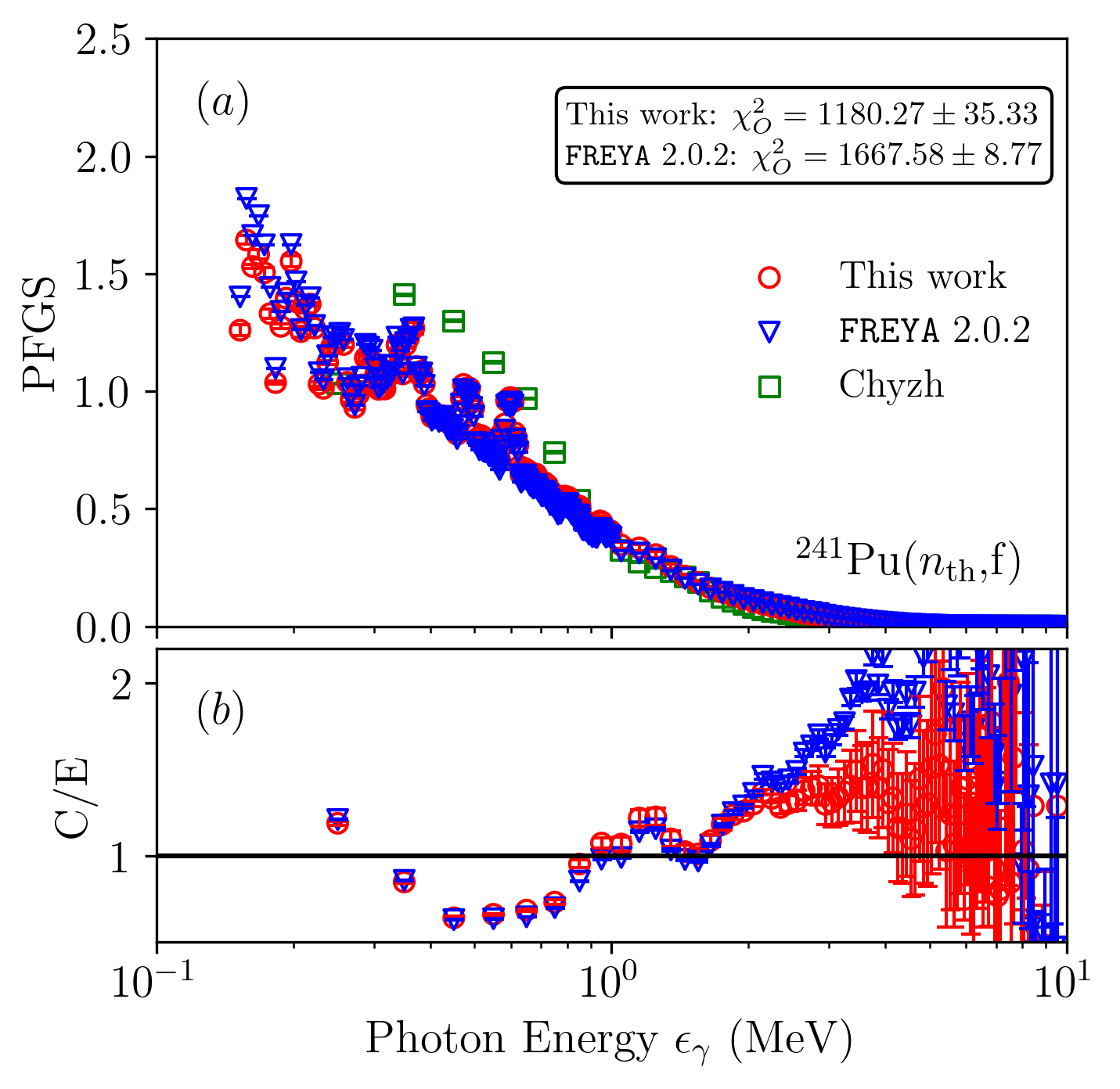}
    \caption{Same as Fig.~\ref{fig:U235nf_PFGS} but for $^{241}$Pu($n_{\rm th}$,f) and experimental data from Ref.~\cite{Chyzh2014}. As discussed in Sec.~\ref{sec:data}, these data were not included in the fit for $^{241}$Pu($n_{\rm th}$,f).}
    \label{fig:Pu241nf_PFGS}
\end{figure}

Figures~\ref{fig:U235nf_PFGS} and \ref{fig:Pu241nf_PFGS} show comparisons to the PFGS data from Ref.~\cite{Oberstedt2013} for $^{235}$U($n_{\rm th}$,f) and from Ref.~\cite{Chyzh2014} for $^{241}$Pu($n_{\rm th}$,f) respectively.  The data from Oberstedt {\it et al.}~\cite{Oberstedt2013} in Fig.~\ref{fig:U235nf_PFGS} are in much finer energy bins than the $\mathtt{FREYA}$ calculation, even when the bin widths of the calculation below 1~MeV were reduced. $\mathtt{FREYA}$ generally reproduces the structures in the PFGS because it connects to the RIPL-3 tables for low energy photon transitions.  Other codes like GEF~\cite{gef2025} that employ only statistical photon emission cannot reproduce the PFGS in this region.

The PFGS structure in the Oberstedt data is not seen in the $^{241}$Pu($n_{\rm th}$,f) Chyzh data~\cite{Chyzh2014} in Fig.~\ref{fig:Pu241nf_PFGS}.  This may be due to lower energy resolution of the detector used for the measurement.  Some of the discrepancy in the comparison may be due to the finer photon energy bins employed in the calculation. The agreement with the data is also poor above $\sim 3$~MeV which may be due in part to statistical uncertainties in the data, not seen in the figure with the $y$ axis shown on a linear scale.

\begin{figure}
    \includegraphics[width=1.0\linewidth]{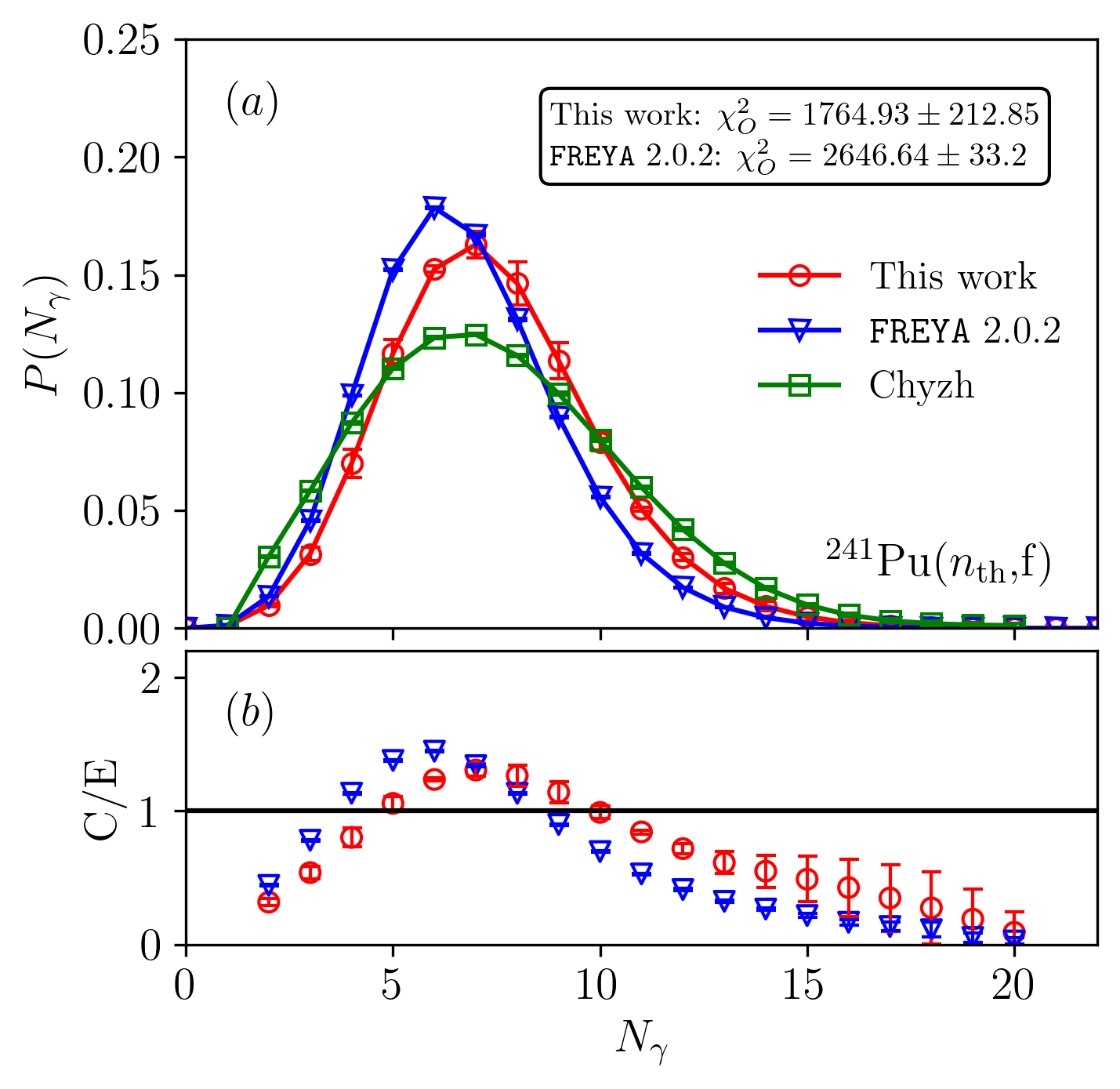}
    \caption{(a) The photon multiplicity distribution for $^{241}$Pu($n_{\rm th}$,f) using parameter values from this work and Ref.~\cite{Verbeke2018} as well as experimental data Ref.~\cite{Chyzh2014}. As discussed in Sec.~\ref{sec:data}, these data were not included in the fit for $^{241}$Pu($n_{\rm th}$,f). (b) Ratio of $\mathtt{FREYA}$ calculations to the experimental results.}
    \label{fig:Pu241nf_P_m}
\end{figure}

Finally, we compare our calculations of the photon multiplicity distribution, $P(N_\gamma)$, to the data from Ref.~\cite{Chyzh2014} for $^{241}$Pu($n_{\rm th}$,f) in Fig.~\ref{fig:Pu241nf_P_m}. As discussed in Sec.~\ref{sec:data}, these data were not included in the fit. $\mathtt{FREYA}$ reproduces the data relatively well, at least in terms of average multiplicity.  The calculations do not reproduce the other moments of the distribution since the shape is much narrower than the data.   Similarly, poor shape agreement with $P(N_\gamma)$ for $^{252}$Cf(sf) was observed in  Ref.~\cite{VanDyke2019}.  In that work, it was noted that there was an uncertainty of $\pm 1$ in the detected photon multiplicity due to multiple scattering.  Thus the $\mathtt{FREYA}$ output was adjusted to account for multiple scattering which broadened the distribution, giving a shape closer to that of the data.  Given that these data were taken by the same group, one might expect that a similar improvement in the comparison to the data might be found here.

Overall, the agreement with the photon data is not as good as that found with the neutron data.  There is the potential for improvement in these comparisons that can be explored in future publications.

\section{Perturbing the genetic algorithm hyperparameters for $^{252}$Cf(sf)}
\label{sec:B}

In our analysis, there are six hyperparameters in the genetic algorithm.  The values of these hyperparameters must be specified: the population size, $p$; the selection size, $k$; the crossover rate, $c$; the mutation rate, $m$; the minimum diversity threshold, $D_{\rm min}$; and the generation number, $g$. In this appendix, we present the $^{252}$Cf(sf) fit results for several perturbations of the hyperparameters from the values we adopt: 
\begin{align}
    (p, k, c, m, D_{\rm min}, g) = (50, 3, 0.80, 1/f, 0.20f, 50) \nonumber
\end{align}
where the number of features is fixed at $f=58$.

The population size, $p$, determines the number of individuals that comprise a generation. While choosing a too small population size can inhibit the search, greater population sizes increase the computational cost of the genetic algorithm. Choosing a value of $p$ is a balance between precision and resource availability. 

The generation number, $g$, sets the number of populations that the genetic algorithm considers. Generally, larger values of $g$ produce better results by providing the algorithm with more iterations to refine its search but this also comes with increased computational demands. Indeed, setting the generation number to be arbitrarily large can yield diminishing returns. 

The number of individuals selected at a time from the population to be considered as parents of the next generation is determined by the selection size, $k$. In effect, this hyperparameter determines the number of individuals in the population that {\it cannot} be selected as parents for the next generation. For $k=3$, for example, the two least fit individuals ({\it i.e.}, those with the two lowest fitness scores) in the population will never be selected as parents, for any combination of individuals, since we only pass on the most fit individual of the size $k$ as a parent. Setting $k$ too large can bias the genetic algorithm by excluding individuals containing desirable characteristics (but not necessarily with high fitness scores), whereas making $k$ too small can delay convergence by keeping undesirable characteristics in the gene pool. 

The crossover and mutation operations are applied to parents in order to create children that populate the next generation. The crossover rate, $c$, determines the frequency at which parents swap genes to make new children. If $c$ is set too high, desirable characteristics can be washed out due to the randomness of the gene swap. However, making $c$ too low can delay convergence by creating future generations too similar to the previous ones. 

In addition, if the mutation rate, $m$, is too high, convergence will be delayed or, in the extreme, the algorithm may never converge.  Thus $m$ is the only non-static hyperparameter in our genetic algorithm: its value is increased when the population diversity, $D$, drops below the minimum diversity threshold, $D_{\rm min}$. The increase in $m$ is proportional to the difference between the minimum diversity threshold and the population diversity, $D_{\rm min}-D$. 

Perhaps a natural question to ask is if there are optimal, or at least equally performing, values for these hyperparameters given the optimization problem. Reference~\cite{Sipper2018} attempted to answer this question by using a meta-genetic algorithm to optimize the hyperparameters of a genetic algorithm employed on 25 different optimization problems. The study totaled nearly $10^5$ runs of the meta-genetic algorithm, each with a population size and generation number that could both be as high as 2,000. Sets of hyperparameters found by the meta-genetic algorithm that yielded solutions of ``good'' quality were recorded for each optimization problem type. In the end, the authors found that comparably good hyperparameter combinations were distributed broadly across the hyperparameter space, indicating a relatively weak sensitivity of the results to the precise hyperparameter values.

\begin{figure}
    \includegraphics[width=1.0\linewidth]{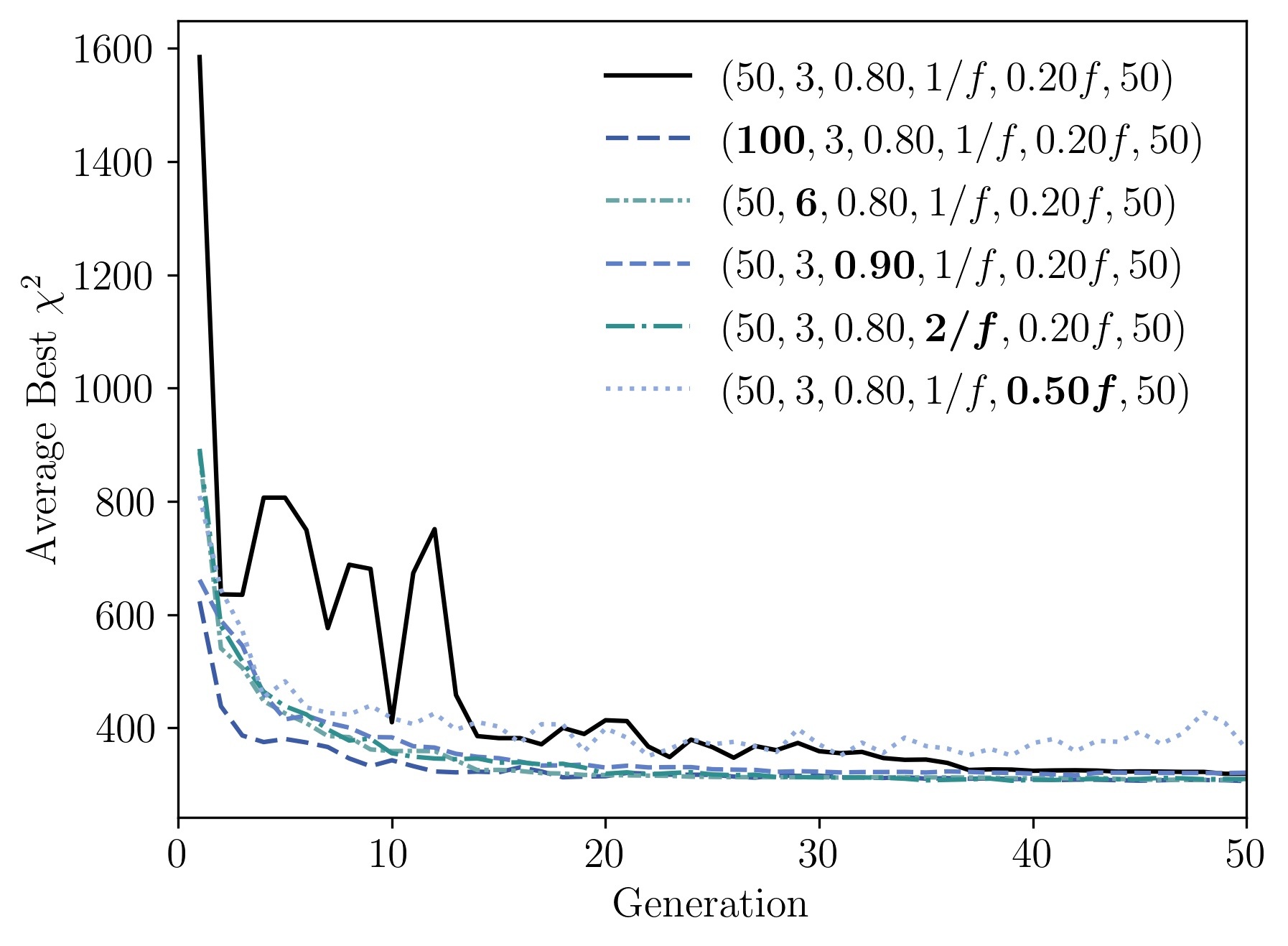}
    \caption{The ensemble average best $\chi^2$ per generation for $^{252}$Cf(sf) using the hyperparameter sets $(p,k,c,m,D_{\rm min},g)$ in Table~\ref{tab:hyperparameters}.}
    \label{fig:convergence}
\end{figure}

We vary the hyperparameters for the $^{252}$Cf(sf) fit, the isotope with the most data available and therefore  placing the greatest constraints on the parameters. Table~\ref{tab:hyperparameters} presents the fit results for different sets of hyperparameters, with each row corresponding to a perturbation of a single hyperparameter from the ``default'' value. Most of the $\mathtt{FREYA}$ parameters agree within uncertainties between ensembles.  Even when there are discrepancies, the parameter values are still similar. Moreover, the minimum $\chi^2$ achieved among these ensembles agree within uncertainties, suggesting that the fits are of similar quality. We do not perturb the generation number, $g$, since our algorithm converged with this number of iterations for all hyperparameter sets considered, see Fig.~\ref{fig:convergence}.  Given the general uniformity of the results across the hyperparameter variations, we therefore employ the ``default'' hyperparameters for both $^{252}$Cf(sf) and all the other systems studied. 

\begin{table*}
\caption{The optimized $\mathtt{FREYA}$ parameters and their corresponding $\chi^2$ for $^{252}$Cf(sf) using different sets of genetic algorithm hyperparameters.}
\label{tab:hyperparameters}
\begin{ruledtabular}
\begin{tabular}{lccccccc}

$(p,k,c,m,D_{\rm min},g)$ & $e_0$ (MeV$^{-1}$) & $x$ & $c$ & $c_S$ & $d$TKE (MeV) & $\chi^2$ \\

\colrule
$(50,3,0.80,1/f,0.20f,50)$ & $ 10.430 \pm 0.496$ & $1.260 \pm 0.006$ & $1.185 \pm 0.021$ & $0.890 \pm 0.064$ & $0.508 \pm 0.293$ & $483.05 \pm 24.63$ \\

\colrule
$({\bf 100},3,0.80,1/f,0.20f,50)$ & $ 11.064 \pm 0.326$ & $1.251 \pm 0.004$ & $1.123 \pm 0.027$ & $0.802 \pm 0.045$ & $0.423 \pm 0.190$ & $483.38 \pm 21.39$ \\

\colrule
$(50,{\bf 6},0.80,1/f,0.20f,50)$ & $ 10.016 \pm 0.182$ & $1.250 \pm 0.002$ & $1.193 \pm 0.013$ & $0.938 \pm 0.048$ & $0.587 \pm 0.115$ & $482.47 \pm 21.20$ \\

\colrule
$(50,3,{\bf 0.90},1/f,0.20f,50)$ & $ 10.437 \pm 0.228$ & $1.233 \pm 0.004$ & $1.149 \pm 0.022$ & $1.001 \pm 0.049$ & $0.247 \pm 0.180$ & $476.48 \pm 25.33$ \\

\colrule
$(50,3,0.80,\bm{2/f},0.20f,50)$ & $ 10.984 \pm 0.191$ & $1.247 \pm 0.004$ & $1.158 \pm 0.019$ & $0.951 \pm 0.047$ & $0.157 \pm 0.132$ & $478.68 \pm 20.50$ \\

\colrule
$(50,3,0.80,1/f,\bm{0.50f},50)$ & $ 10.927 \pm 0.122$ & $1.234 \pm 0.007$ & $1.180 \pm 0.017$ & $0.847 \pm 0.055$ & $0.426 \pm 0.099$ & $481.73 \pm 21.07$ \\

\end{tabular}
\end{ruledtabular}
\end{table*}

\bibliography{references}

\end{document}